\documentclass[fleqn,usenatbib]{mnras}

\usepackage{newtxtext,newtxmath}

\usepackage[T1]{fontenc}

\DeclareRobustCommand{\VAN}[3]{#2}
\let\VANthebibliography\thebibliography
\def\thebibliography{\DeclareRobustCommand{\VAN}[3]{##3}\VANthebibliography}

\usepackage{graphicx}	
\usepackage{amsmath}	
\usepackage{hyperref}
\usepackage[nopatch]{microtype}
\usepackage{booktabs}
\usepackage{xcolor}
\usepackage{tikz}
\usepackage{tabularx}
\usepackage{graphicx}	
\usepackage{amsmath}	
\usepackage{multirow}
\usepackage{longtable}
\usepackage{array}
\usepackage[export]{adjustbox}
\usepackage{rotating}
\usepackage{collcell}   
\usepackage{multirow}
\usepackage{longtable}
\usepackage{array}

\def\farcs{%
 \mbox{%
  \kern  0.13ex.%
  \kern -0.95ex\arcsec%
  \kern -0.1ex%
 }%
}%

\newcolumntype{P}[1]{>{\centering\arraybackslash}p{#1}}
\newcolumntype{C}[1]{>{\centering\arraybackslash}m{#1}}

\newcommand*{\MinNumber}{-1.0}%
\newcommand*{\MidNumber}{0.0} %
\newcommand*{\MaxNumber}{1.0}%
\definecolor{amber}{rgb}{1.0, 0.49, 0.0}
\definecolor{aqua}{rgb}{0.0, 1.0, 1.0}
\def\ifempty#1{\def\temp{#1}\ifx\temp\empty}
\newcommand{\ApplyGradient}[1]{%
    \ifempty{#1}
        #1
    \else
        \ifdim #1 pt > \MidNumber pt
            \pgfmathsetmacro{\PercentColor}%
                {max(min(100.0*(#1-\MidNumber)/(\MaxNumber-\MidNumber),100.0),0.00)}%
            \hspace{-0.95mm}\colorbox{amber!\PercentColor!white}{#1}
        \else
            \pgfmathsetmacro{\PercentColor}%
                {max(min(100.0*(\MidNumber-#1)/(\MidNumber-\MinNumber),100.0),0.00)}%
            \hspace{-0.95mm}\colorbox{aqua!\PercentColor!white}{#1}
        \fi
    \fi
}
\newcolumntype{Z}{>{\collectcell\ApplyGradient}c<{\endcollectcell}}
\title[Milky Way Planetary Nebulae: Properties \& Distribution]{Analysis of Planetary Nebulae in the Milky Way: Physical Properties, Chemical Abundances, and Galactic Distributions}

\author[N. Erzincan et al.]{
	N. Erzincan$^{1}$\thanks{E-mail: nerzincan@student.cu.edu.tr},
	N. Aksaker$^{1,2}$,
    A. Akyuz$^{1,3}$
    And
    Q. Parker$^{4}$
	\\
	$^{1}$Space Science and Solar Energy Research and Application Center (UZAYMER), University of Çukurova, 01330, Adana, Türkiye\\
	$^{2}$Adana Organised Industrial Zones Vocational School of Technical Science, Çukurova University, 01410, Adana, Türkiye\\
	$^{3}$Department of Physics, University of Çukurova, 01330, Adana, Türkiye\\
    $^{4}$The Laboratory for Space Research, The University of Hong Kong\\
}

\date{Accepted XXX. Received YYY; in original form ZZZ}

\pubyear{\the\year{}}

\begin{document}
\label{firstpage}
\pagerange{\pageref{firstpage}--\pageref{lastpage}}
\maketitle

\begin{abstract}
In this study, we investigate the physical and chemical properties of planetary nebulae (PNe) from the Milky Way Galaxy using the largest number of sources to date, with 1,449 True PNe from the HASH database. Among the Galactic components—thin disk, thick disk, halo, and bulge—most PNe are concentrated in the Galactic disk, with a median angular size of 12 arcseconds (0.45 pc), while halo PNe tend to have larger sizes. Physical parameters of whole PNe, extinction coefficent c($H_\beta$), electron temperature (T$_e$), and density (N$_e$) show Gaussian-like distributions with medians of 1.5, 9,900 K, and 1,1200 cm$^{-3}$, respectively. The abundances of He, N, O, Ne, S, Cl, and Ar in PNe show Gaussian distributions with slight variations across Galactic components. PNe located in thin disk exhibit higher abundances, except for O and Ne, while PNe in halo have the lowest values for all elements. Strong correlations between elements, particularly Sulphur vs. Nitrogen (r=0.87), were identified using statistical tests. Comparisons with previous studies reveal variations (< 2 dex.) in abundance ratios, particularly in halo PNe. We also present the first detailed database in the literature, providing $\sim$7,200 abundance values for these elements, derived from $\sim$16,500 emission line measurements, to support the testing and development of theoretical models.
\end{abstract}




\section{Introduction}

Planetary nebulae (PNe) are objects that contribute to the enrichment of the Interstellar Medium (ISM) during the transition from  the Asymptotic Giant Branch (AGB) stage to the white dwarf phase, with masses ranging from about 0.8 to 8 M$_{\odot}$ \citep{1991A&A...243..478Z}.
Despite their relatively short lifetime of 3-7 x 10$^4$ years, the study of PNe abundances offers valuable insights into the composition of the ISM \citep[hereafter KH22]{Kwitter2022}, \citep{2017IAUS..323..339M}. Elements such as helium (He), carbon (C), nitrogen (N), oxygen (O), neon (Ne), sulfur (S), chlorine (Cl), and argon (Ar) produce prominent emission lines in PNe, resulting from nucleosynthesis processes in the progenitor star. Accurate measurements of these spectral lines enable the calculation of abundances and physical conditions, providing a fundamental understanding of the physical and chemical properties of PNe and their evolutionary history (KH22, \citealp{2022Galax..10...32P}).

In the review paper by KH22, various observed properties of PNe were discussed, including determining element abundances, comparisons with theoretical predictions, and the presence of associated dust and molecules. The paper also examined key aspects such as distances, binarity, morphologies, and evolution of PNe. In their study, it is emphasized that the He and N abundances of PNe in the bulge and disk are above solar abundances, O and Ar are near solar, while S is below. Additionally, the Ne abundance in the disk component is 30$\%$ higher than in the bulge region, while the Cl abundance in the bulge region is twice as high as in the disk. Moreover, in the halo, O, Ne, S, and Ar abundances are below solar abundances, while He, C, and N are above solar levels.

The chemical abundances of PNe located in the Galactic bulge were analyzed by \cite{2024MNRAS.527.6363T} (hereafter Tan24) using high-resolution spectroscopic data.
Their abundance compilation shows overall higher elements (N, O, Ne, S, Cl and Ar) abundances than solar, differing from the general pattern in the literature. The abundance studies of \cite{chiappini2009} on PNe in our Galaxy's bulge and disk also discussed that, at the bulge metallicity, O and Ne abundances are close to the interstellar medium values at the time of PN progenitor formation, making them tracers of the bulge chemical evolution, similar to S and Ar.

\cite{2024ApJ...972..130S} (hereafter S24) investigated the abundances of C, N, O, Fe (iron), and S in PNe within the Galactic disk, suggesting that these nebulae are descendants of red giant stars. The study revealed significant Fe depletion in PNe, with most of the Fe trapped in dust grains, along with slight S depletion. Additionally, evidence of C and N enrichment was found in the progenitor stars, attributed to hot-bottom burning. Additionally, many studies such as those \citealp{2001AmSci..89..506C, 2004ApJ...617.1115D, 2011EP&S...63.1067K, Pagomenos2018}, have examined elemental abundances in the Milky Way, focusing on aspects such as metallicity gradients, chemical evolution, and the composition of various Galactic components.

High-resolution narrow-band imaging and wide-field surveys in optical, infrared, and radio wavelengths allow a reevaluation of the identities and morphologies of PNe in existing catalogs. This effort will contribute to the Hong Kong/Australian Astronomical Observatory/Strasbourg Observatory H$\alpha$ Planetary Nebula\footnote{\url{http://hashpn.space}} (HASH) database for PNe research by integrating new data with existing spectroscopy \citep{parker2016}. The HASH database is currently contains 11,352 objects, both Galactic and extragalactic. 
Detailed information about HASH is given in Section \ref{sec:HASH}. In this work, we aim to perform a comprehensive abundance study of the Milky Way using the spectra of 2,591 PNe available in the HASH database. By utilizing distances from the Gaia EDR3 catalog, based on the central stars of planetary nebulae (CSPNe), we determined the positions of these PNe within the bulge, thin disk, thick disk, and halo components of the Galaxy, and investigated the abundance gradients among these components. After downloading the spectral data of PNe from the HASH database, we calculated emission line fluxes and applied reddening corrections using the extinction coefficient. Based on these corrected fluxes, the physical properties and chemical abundances of each PNe were analyzed.

Our study is organized as follows: Section \ref{sec:obs} presents the analysis of the spectral data. Section \ref{sec:pcc} details the physical and chemical conditions derived from the emission line ratios. The results and discussion are provided in Section \ref{sec:res_dis}.

\section{Data analysis}
\label{sec:obs}

\begin{figure}
    \centering
    \includegraphics[width=0.95\linewidth]{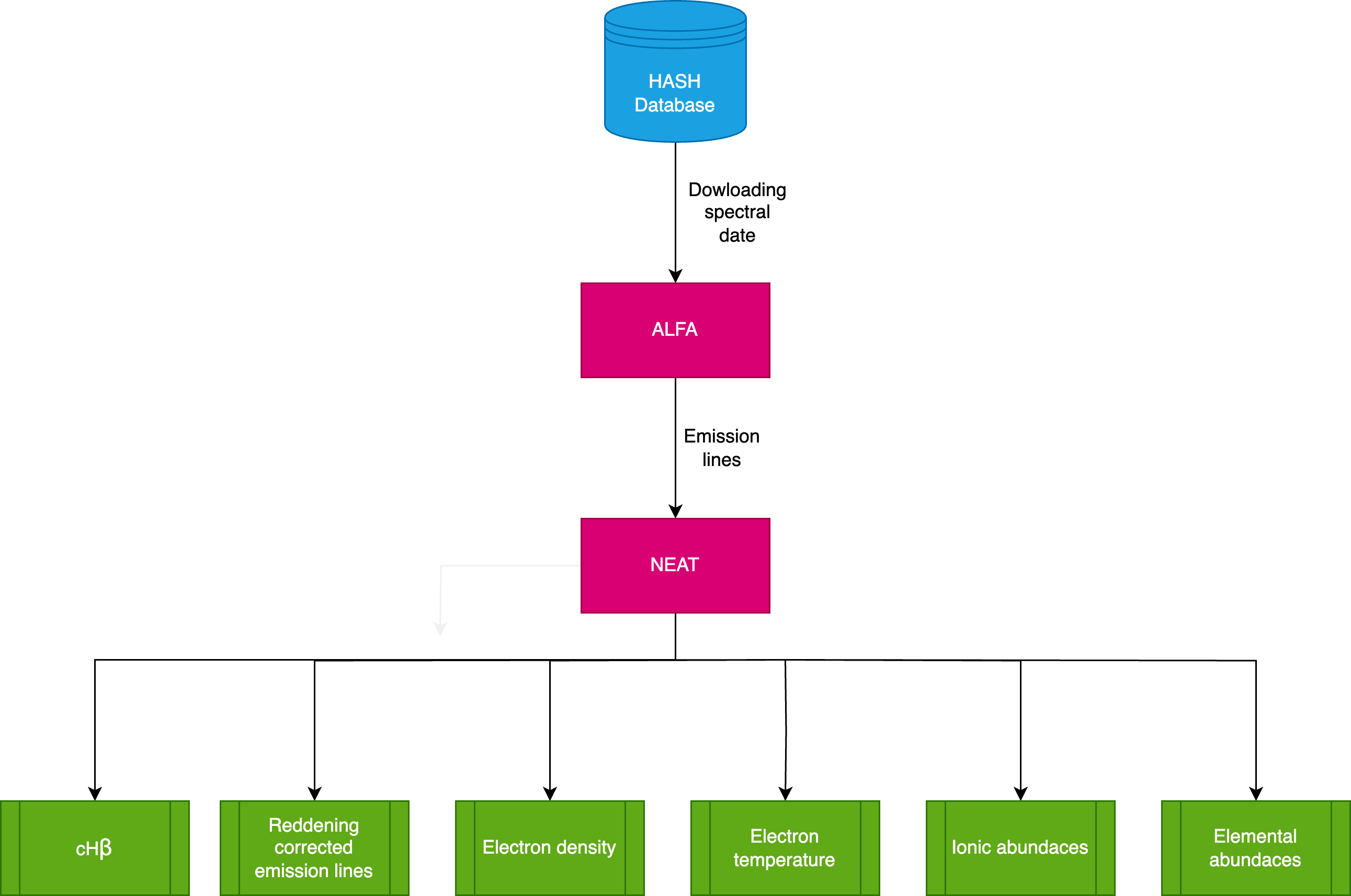}
    \caption{The flowchart of the study.}
    \label{F:fchart}
\end{figure}

Our study consists of three main stages: obtaining the spectra of the PNe from the HASH database, measuring their emission lines, and determining the physical and chemical properties. The general flowchart of the study is shown in Fig. \ref{F:fchart}.

\subsection{HASH Database}
\label{sec:HASH}

The HASH database \citep{parker2016, 2017IAUS..323.....L} is a comprehensive collection of photometric and spectral data for different objects. It encompasses spectral data from 197 different catalogs and 145 independent studies compiled since March 5, 1970. The database classifies objects into 45 distinct types, including T: PNe, PNe candidates (designated as L: Likely PNe, P: Possible PNe, c: New Candidates) N: Not PNe, Supernova Remnants (SNR), H II, galaxies (G), and others. Additionally, it provides information on up to 36 characteristics for most objects, including images across UV to radio wavelengths, as well as details on names, statuses, coordinates, catalog references, physical dimensions, and subclasses. The database is continuously updated with the outputs of the new studies. A sample of the information used in our research is presented in Table \ref{T:HASH} and the complete catalog is available in the supplementary files referenced in Section \ref{sec:Supl}.

\begin{figure}
    \centering
    \includegraphics[width=0.95\linewidth]{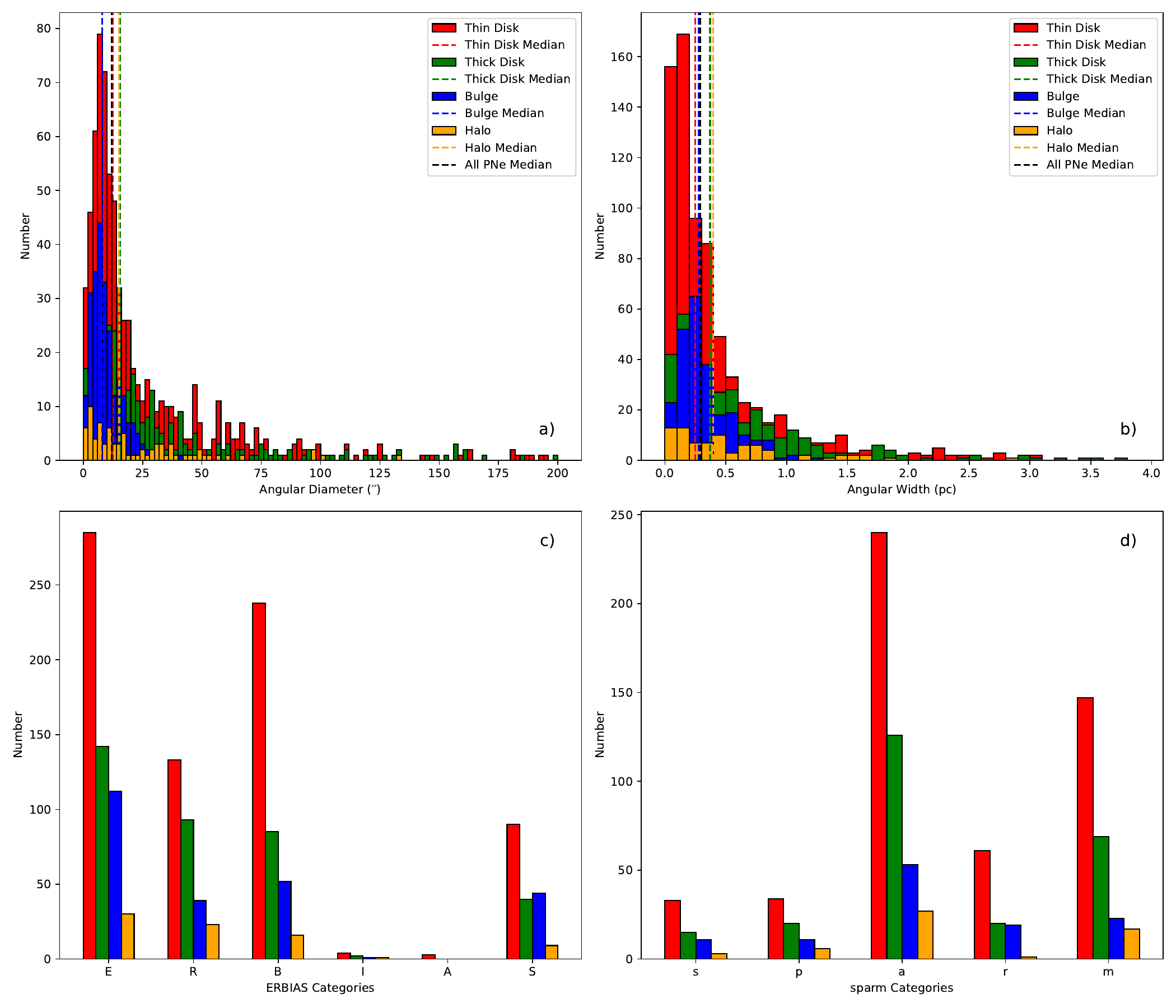}
    \caption{ Histograms of angular sizes of PNe in arcseconds (a) and parsecs (b), and morphological classifications based on the ERBIAS (c) and SPARM (d) schemes. The data are sourced from the HASH database \protect\citep{2023MNRAS.519.1049T}. Colors indicate the Galactic component associated with each PN: bulge (blue), thin disk (red), thick disk (green), and halo (orange). The black dashed line shows the overall mean value, while the colored dashed lines indicate the median values for each Galactic component.
    }
    \label{F:Size_Morpho}
\end{figure}

Here, we examined 2,591 True PNe (hereafter referred to as PNe) cataloged in the HASH database until December 8, 2023. These cataloged PNe have multiple spectral datasets from 45 different studies, with a total of 4,925 spectra available. Among these, PN G357.7-04.8, which has the most spectral data in the database, contains 43 different datasets in the optical region. All spectra were downloaded from the HASH database in FITS format.

The angular dimensions of PNe sources in the HASH catalog contain morphological information organized by main classes and subclasses. These main classes are defined as ERBIAS: Elliptical (oval), round, bipolar, irregular, asymmetric, or quasi-stellar (point source). In addition, subclasses are specified for features such as has resolved internal structure (s), exhibiting point symmetry (p), one-sided enhancement/asymmetry (a), has a well-defined ring structure or annulus (r) or having multiple shells or external structure (m). These classifications have been used according to their presence in the components of Galaxy, based on the scheme presented by \cite{2006MNRAS.373...79P}, and are shown in Fig. \ref{F:Size_Morpho}. The components of the Galaxy are as follows: bulge (blue), thin disk (red), thick disk (green) and halo (orange). This notation is used throughout the study.

\subsection{Distance}
\label{sec:Distance}

\begin{figure}
\begin{center}
\includegraphics[width=0.95\linewidth]{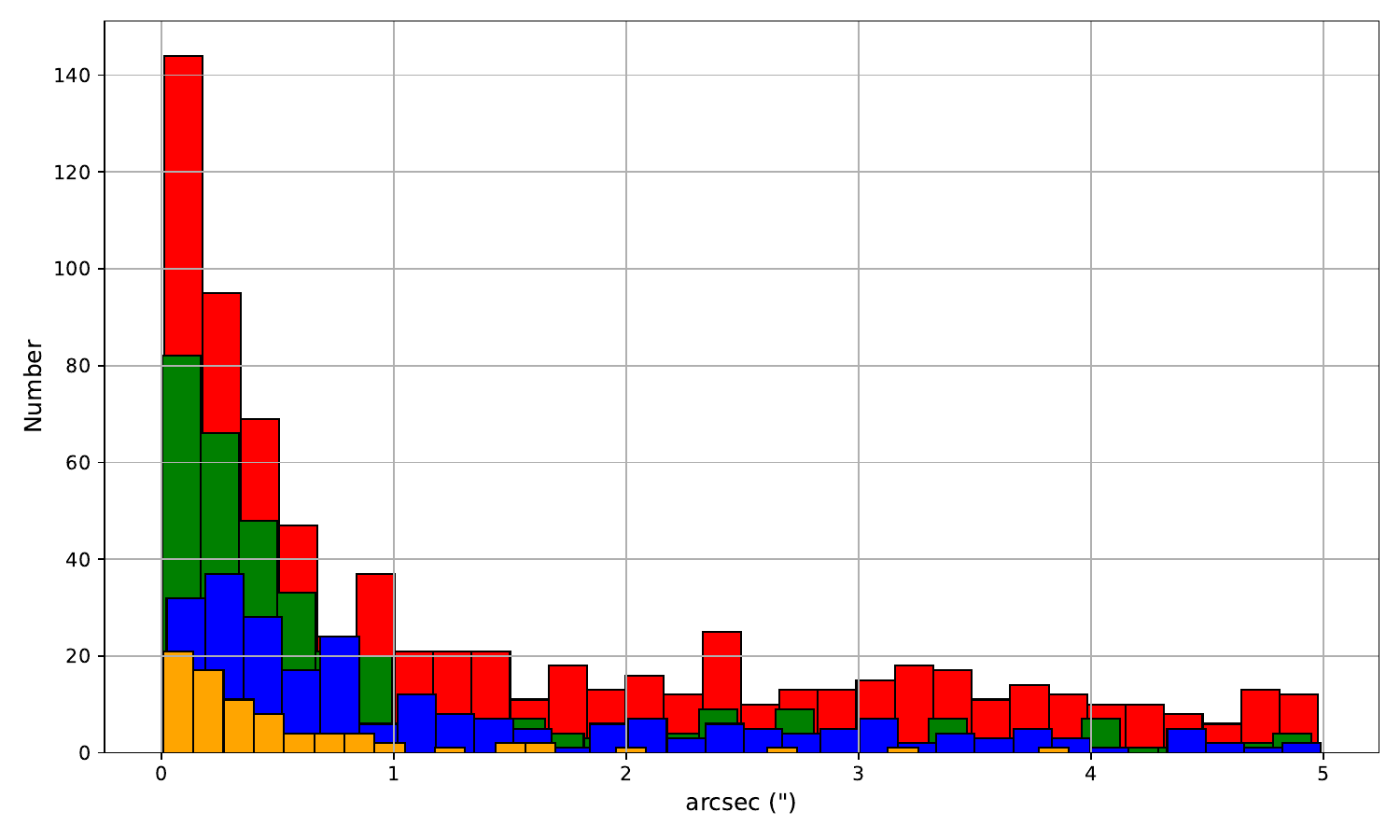}
\caption{Histogram displaying the angular distances between the HASH coordinates and the matched Gaia EDR3 source coordinates for each PNe. The Galactic components are color-coded as follows: bulge (blue), thin disk (red), thick disk (green), and halo (orange).}
\label{F:r_dist}
\end{center}
\end{figure}

\begin{figure}
\begin{center}
\includegraphics[width=0.95\linewidth]{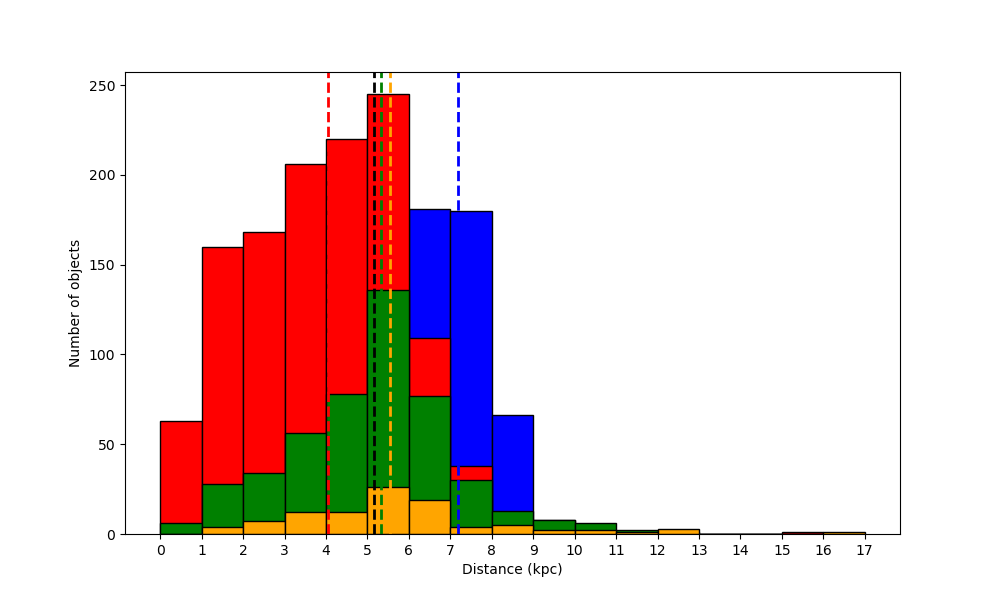}
\caption{The histogram of distances of  PNe from the Gaia EDR3 database. The Galactic components are color-coded as follows: bulge (blue), thin disk (red), thick disk (green), and halo (orange).}
\label{F:dist}
\end{center}
\end{figure}

Distance information for PNe can be derived from the properties of the CSPNe and/or the nebulae (KH22). In this study, we used the Galactic coordinates of the PNe listed in the HASH database to query the Gaia EDR3 catalogue\footnote{\url{https://vizier.cds.unistra.fr/viz-bin/VizieR-3?-source=I/352/gedr3dis}} \citep{2021AJ....161..147B}(hereafter BJ21). For each PN, we selected the closest Gaia EDR3 source within a 5-arcsecond search radius to ensure reliable matching. Fig. \ref{F:r_dist} also presents a histogram of the angular separation between each PN and its matched Gaia EDR3 source, based on Galactic coordinates. The \textit{Rgeo} values used in this study represent the median of the geometric distance estimates provided by BJ21, and are given in kiloparsecs (kpc) in the leftmost column of Table \ref{T:HASH}. The distribution of \textit{Rgeo} is shown separately in Fig. \ref{F:dist}.

\begin{figure}
\begin{center}
\includegraphics[width=0.95\linewidth]{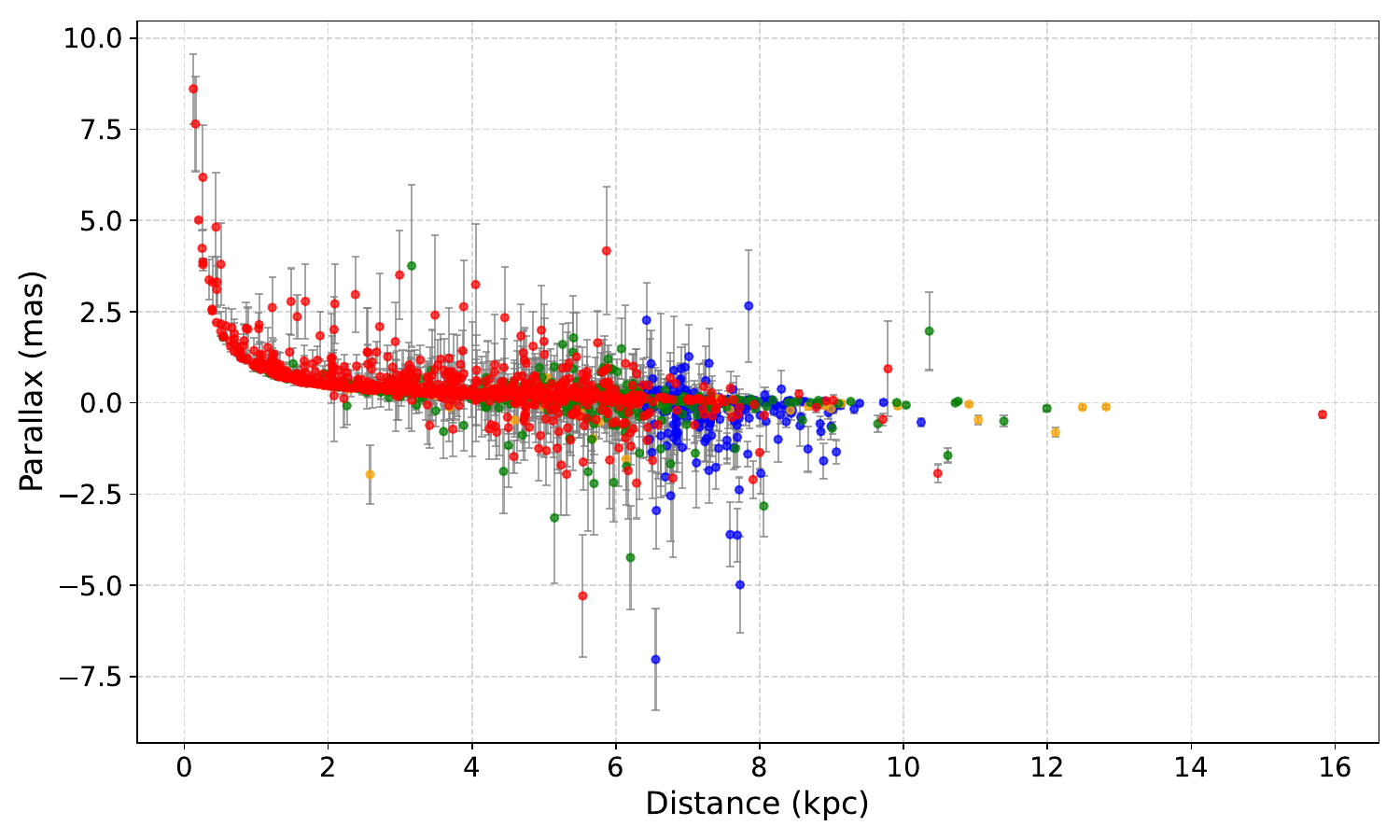}
\caption{Scatter plot illustrating the relationship between Gaia EDR3 geometric distances (\textit{Rgeo}) and the corresponding parallax values with their uncertainties, shown as vertical error bars. The Galactic components are color-coded as follows: bulge (blue), thin disk (red), thick disk (green), and halo (orange).}
\label{F:r_parallax}
\end{center}
\end{figure}

We retrieved parallax and parallax uncertainty values from the Gaia EDR3 catalogue\footnote{\url{https://vizier.cds.unistra.fr/viz-bin/VizieR-3?-source=I/350/gaiaedr3}} for further analysis (BJ21). It is well established that parallax measurement errors increase with distance, which significantly reduces the reliability of distance estimates for more remote sources (BJ21). As illustrated in Fig.~\ref{F:r_parallax}, parallax values remain above 1 mas at distances less than 1 kpc but decline steadily with increasing distance, reaching approximately 0.16 mas at 5–6 kpc and around 0.08 mas at 6–7 kpc.

The distances determined by Gaia EDR3 from the center coordinates of PNe may not accurately represent the positions of the central stars. This situation may lead to deviations in the derived distribution of Galactic components. To evaluate this deviation, we used the dataset from \cite{2021A&A...656A.110C} (hereinafter referred to as C21), which focuses on the central stars of Galactic PNe. C21 provides high-precision astrometric measurements for CSPNe and compact PNe using Gaia EDR3 and identifies a total of 2,117 sources. All sources used in our study are consistent with those defined as True PNe in C21, ensuring a high level of reliability in the comparison. Of the sources in our study, 1,108 (76.5\%) match the C21 catalog. Among these, 84.5\% (936) correspond to the same PNe identified in both studies. The remaining 15.5\% (172) lack Gaia parallax information in the C21 catalog, whereas our dataset includes parallax values and their associated uncertainties for all PNe considered in this study.


The boundaries of the Galaxy's components are poorly defined and remain a subject of debate. In the study by \cite{yan2019}, the boundaries of the thin and thick disks were reported as 200-369 pc and 600-1,000 pc, respectively. However, \cite{ak2021galaksi} defined these components as 250-400 pc and 600-1,500 pc, respectively. Furthermore, \cite{2016PASA...33...25Z} and \cite{ak2021galaksi} highlighted that the bulge region may have a radius of 2.5-3 kpc and 1.5-4.5 kpc, respectively, considering its bar-like structure. They also indicated that the halo region of the Galaxy could extend from 3 to 100 kpc in length \citep{ak2021galaksi}. Additionally, \cite{2014ApJ...783..130R} determined the distance of the Sun from the Galactic center as R$_0$ = 8.34 $\pm$ 0.16 kpc in their study.

\begin{figure}
\begin{center}
\includegraphics[width=0.95\linewidth]{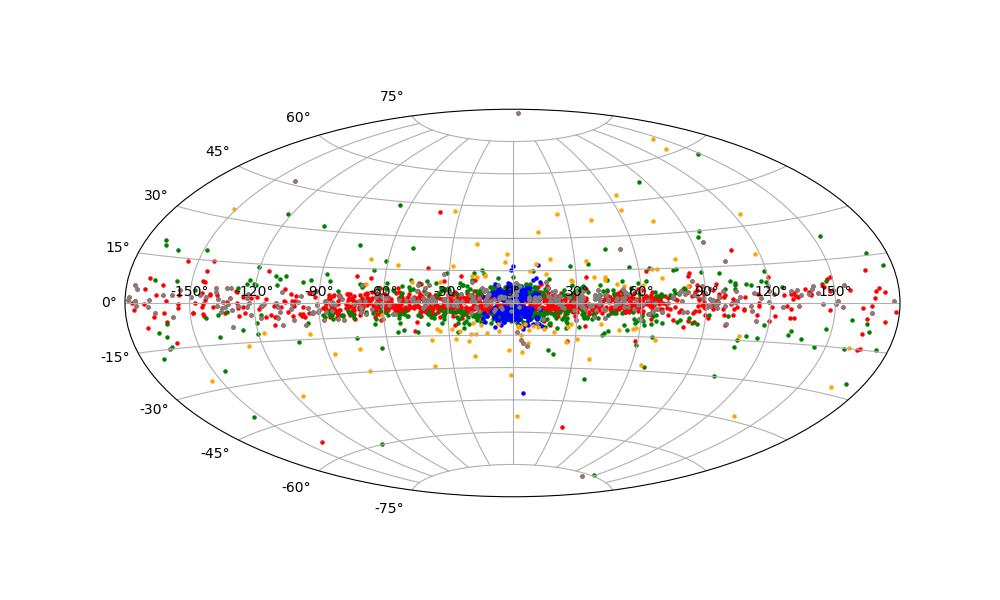}
\caption{True PNe from the HASH database are displayed in an Aitoff projection. Colors indicate PNe in different Galactic components: blue (bulge), red (thin disk), green (thick disk), orange (halo), and gray (no distance information). This color scheme is applied consistently throughout the study.} 
\label{F:Aitoff_sky}
\end{center}
\end{figure}

In this study, we considered a thin disk up to 400 pc from the Galactic plane, a thick disk between 400-1,000 pc, and a spherical bulge region with a radius of 2 kpc at a distance of 8.34 kpc from the Sun. Objects outside these boundaries were classified as halo sources. Taking into account the PNe distances and Galactic longitude information, we used the small angle formula to determine the distance from the Galactic disk. As a result, for 2,591 PNe, the number of sources in the bulge region, thin disk, thick disk, halo, and unknown distances are 437, 1,222, 475, 98, and 359, respectively. The distribution of the PNe in the Galaxy is shown in Fig. \ref{F:Aitoff_sky}. The unknown distances of the PNe are not used in this study.

\subsection{Emission Line Measurements}

Spectral data analysis of PNe is essential for understanding their emission mechanisms. We utilized the automatic line fitting Algorithm ({\scshape alfa}) (version: 2.21\footnote{\url{https://nebulousresearch.org/codes/alfa/}}, \citealp{2016MNRAS.456.3774W}) code, which is a leading software tool for measuring emission lines. {\scshape alfa} measures the fluxes of these lines through a three-step process. First, it extracts the continuum of the spectrum, then, followed by determining the resolution of the spectrum and the velocity of the source. In the last step, the observed spectrum is divided into windows and then they are matched with synthetic lines and observed emission lines. {\scshape alfa} identifies lines likely present in PN spectra. In our study, the {\scshape alfa} code was run with a 5-sigma threshold, yielding a median signal-to-noise ratio of 9.4 for the selected emission lines. This indicates a satisfactory level of precision in the flux measurements, with a typical relative uncertainty around 11\%, which ensures the reliability of the derived emission line parameters. The resultant flux values and errors of the PNe emission lines were obtained by scaling to $H_\beta$=100.

In this study, a total of 4,925 spectra were analyzed using the {\scshape alfa} code. Given the extensive research on PNe, specific selection criteria were applied to refine the sample: (a) spectra containing emission lines in both the blue and red regions, (b) the spectrum of each PN with the highest signal-to-noise ratio for H$\alpha$ at 6562.77 \text{\AA}, and (c) available \textit{Rgeo} data. As a result, a total of 1,449 PNe spectra met the selection criteria and were included in the subsequent analyzes. The spectral coverage was also evaluated, with the median starting wavelength of the selected sources being 4003.58 \text{\AA} and the median ending wavelength 7329.67 \text{\AA}. In total, 437 distinct emission lines were identified across the entire PNe sample. The distribution of these PNe in components of the Galaxy is as follows; 248 in the bulge, 756 in the thin disk, 365 in the thick disk, and 80 in the halo. Table \ref{T:DFlux}, lists the emission lines and flux measurements required to determine the properties for the PNe sample described in Section \ref{sec:pcc}. In addition, all emission lines and flux measurements for the whole PNe are given in Section \ref{sec:Supl}.

\subsection{Diagnostic Diagrams}

Diffuse objects like H II, SNR, and PNe often exhibit similar spectral features, making them challenging to distinguish. Typically, line ratios and diagnostic tools are used to differentiate these sources \citep{2010PASA...27..129F,2022FrASS...9.5287P}. Although the sources used here are classified as PNe, we utilized this large sample to confirm their classification with diagnostic techniques, presenting them in the literature for the first time.

\begin{figure}
\begin{center}
\includegraphics[angle=0,scale=0.33]{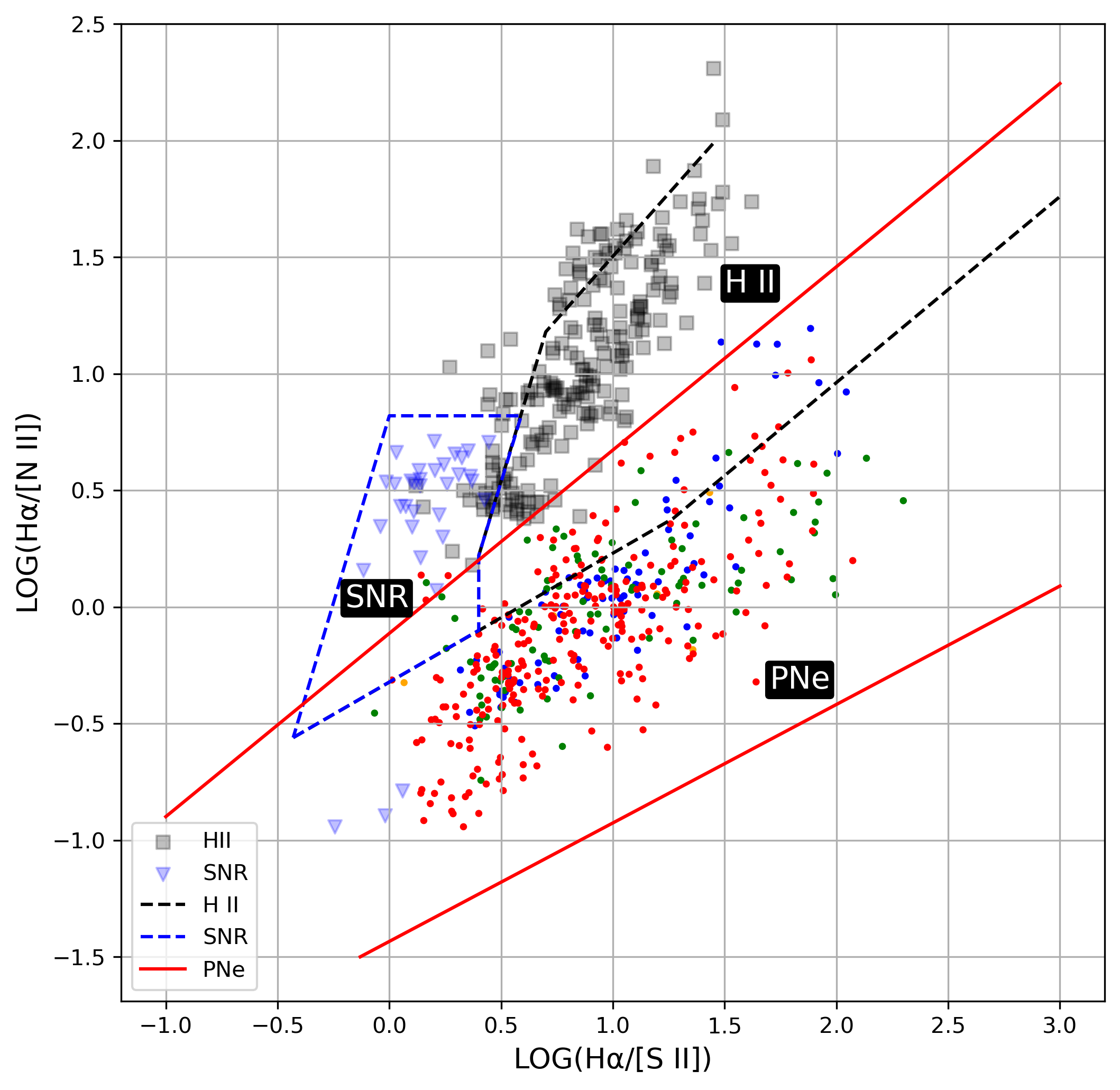}
\includegraphics[angle=0,scale=0.33]{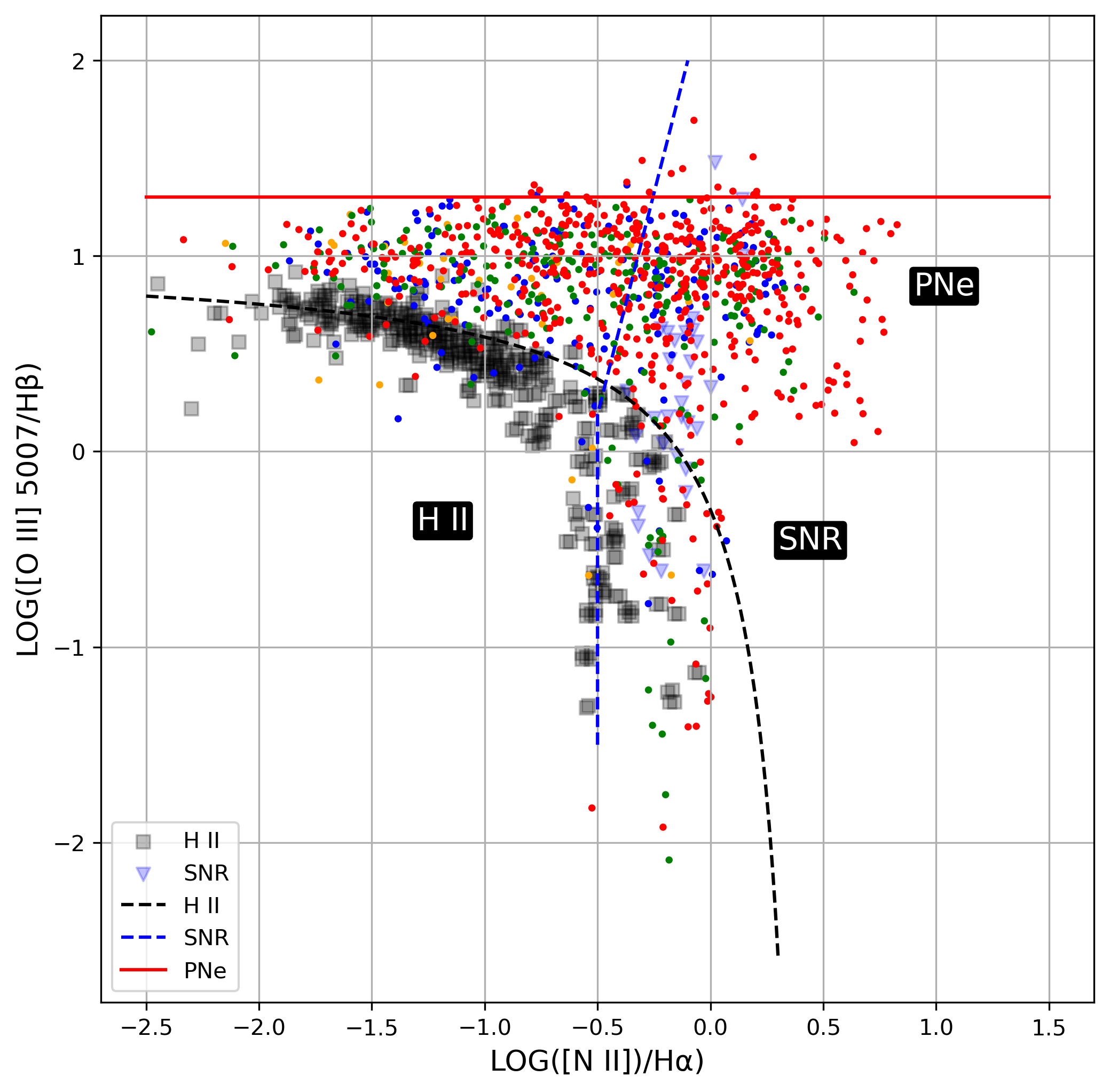}
\includegraphics[angle=0,scale=0.33]{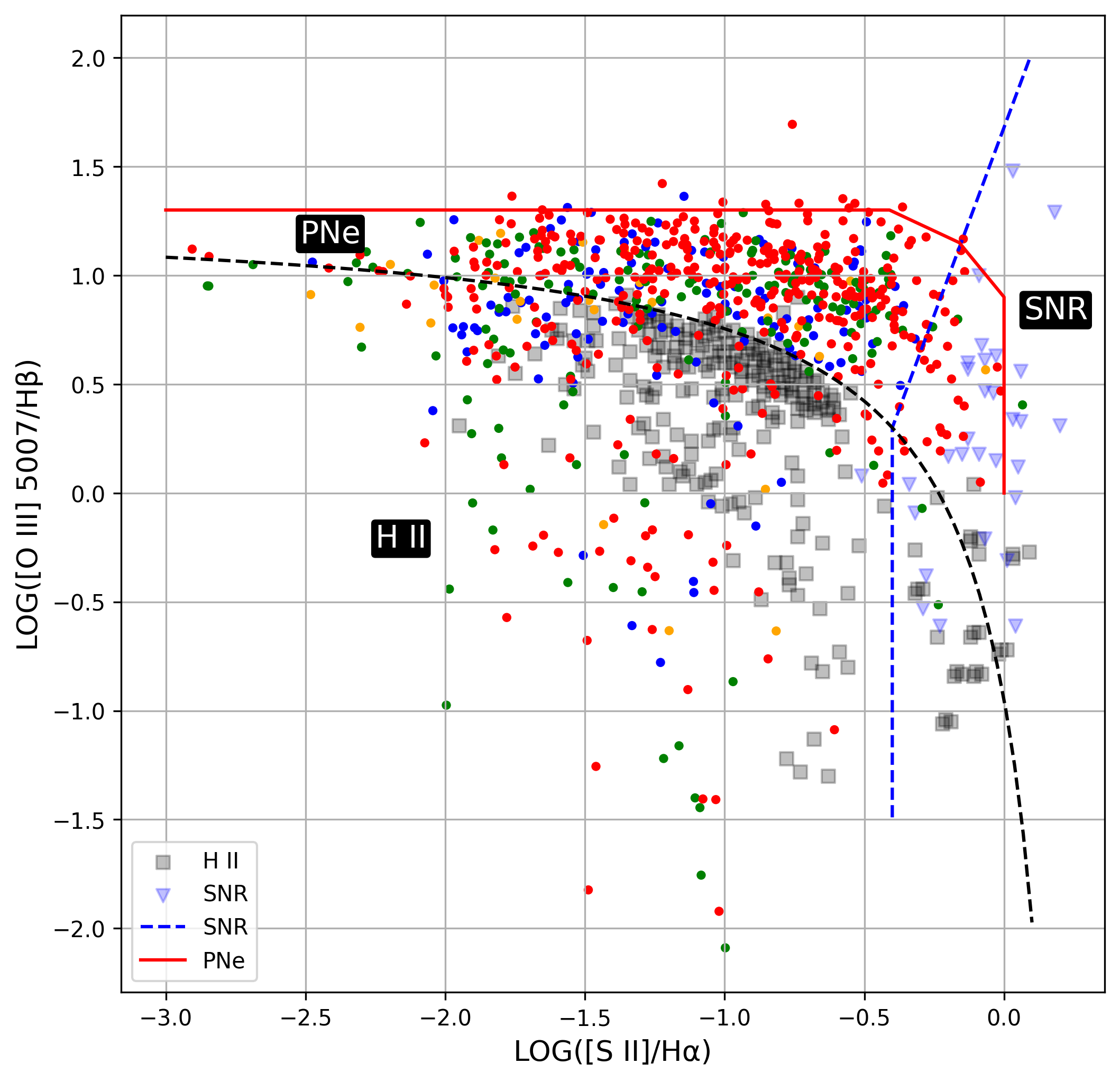} 
\caption{SMB and BPT (a) and BPT (b) diagnostic diagrams were constructed using PNe, H II regions (\protect\citealp{2007MNRAS.381.1719V}), and SNRs (\protect\citealp{2008MNRAS.383.1175P}). The H II region is bounded by black dashed lines, while the Galactic PNe are confined between the red solid lines. SNRs are enclosed within the blue dashed lines. These restricted regions are marked by \protect\cite{2013MNRAS.431..279S}.}
\label{F:smb_bpt}
\end{center}
\end{figure}

The diagnostic diagrams of Sabbadin-Minello-Bianchini (SMB) \citep{1977A&A....60..147S} and Baldwin-Phillips-Terlevich (BPTa and BPTb) \citep{1981PASP...93....5B} are constructed using emission line ratios. For PNe used in these diagrams, the emission lines are $H_\beta$ $\lambda$4861, [O III] $\lambda$5007, $H_\alpha$ $\lambda$6563, double [N II] $\lambda$$\lambda$6548-6584 and double [S II] $\lambda$$\lambda$6716-66731. The line ratios Log(F(X)/[N II]) and Log(F(X)/[S II]) used in the SMB and BPT plots represent the sum of the [N II] and [S II] double emission fluxes. The constructed SMB and BPT diagnostic diagrams are given in Fig. \ref{F:smb_bpt}.

\section{PNe Properties}
\label{sec:pcc}

The interstellar gas and dust extinction in the Galaxy affects the spectral data for PNe. These data are a valuable tool for calculating the H$_\beta$ extinction coefficient, c(H$_\beta$), which determines the interstellar reddening around these objects \citep{2016MNRAS.455.1459F}. The physical and chemical properties of PNe calculated from the reddening-corrected emission line fluxes and line ratios are obtained from the Nebular Empirical Analysis Tool ({\scshape neat}) code (v2.3\footnote{\url{https://github.com/rwesson/NEAT)}}, \citealp{2012MNRAS.422.3516W}), which was used in our study. {\scshape neat} calculates ionic and elemental abundances while also propagating the statistical uncertainties from the input emission line fluxes through all derived parameters. This approach provides a consistent estimate of the uncertainties in the results. However, these uncertainties represent lower limits, since only statistical errors are considered in propagation, excluding potential systematic effects such as flux calibration inaccuracies or line blending.

\subsection{Physical Properties}

\begin{figure}
    \centering
    \includegraphics[width=0.95\linewidth]{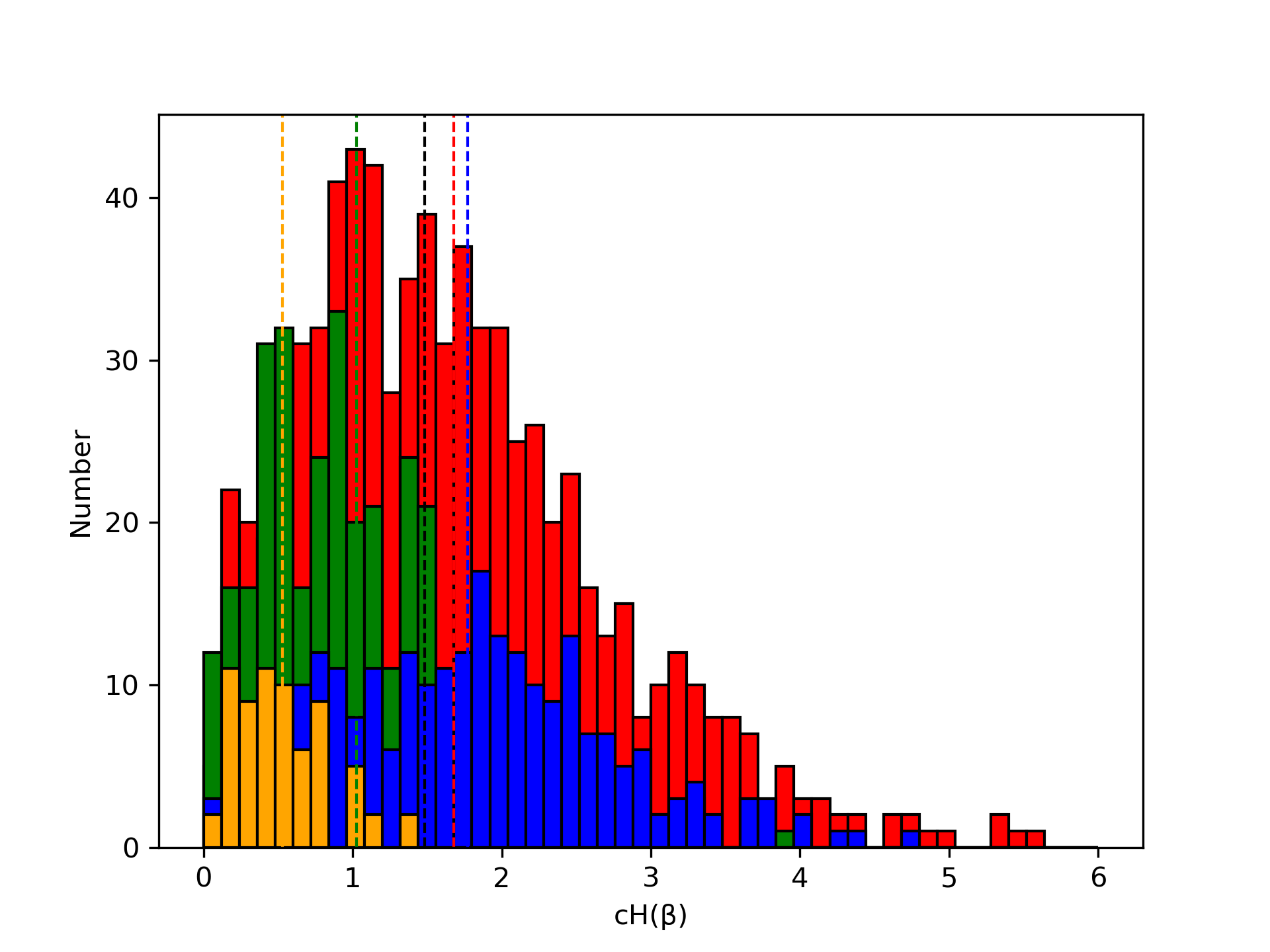}
    \caption{The histogram of the c(H$_{\beta}$) calculated from the line ratio of H$_{\alpha}$ / H$_{\beta}$ in PN spectra. The Galactic components are color-coded as follows: bulge (blue), thin disk (red), thick disk (green), and halo (orange). The dashed lines represent the median values of elemental abundances for the respective Galactic components, while the black line represents the median of all components.}
    \label{f:cbeta}
\end{figure}

Using reddening-corrected emission fluxes provides accurate insight into the surrounding medium. These emission fluxes in the {\scshape neat} code are based on the method described by \cite{1990ApJS...72..163F}. The reddening correction in {\scshape neat} is applied using the line ratios of H$\alpha$ / $H_{\beta}$, $H_{\gamma}$ / $H_{\beta}$, or $H_{\delta}$ / $H_{\beta}$. 
The histogram of the extinction coefficient c($H_\beta$) derived from this $H_{\alpha}$ and $H_{\beta}$ line ratios, commonly used in PN studies. The c ($H_\beta$) values (except for extinction coefficients such as c($H_\gamma$) and c($H_\delta$) were chosen due to the large sample size displayed in Fig. \ref{f:cbeta}.

\begin{figure}
    \centering
    \includegraphics[width=0.95\linewidth]{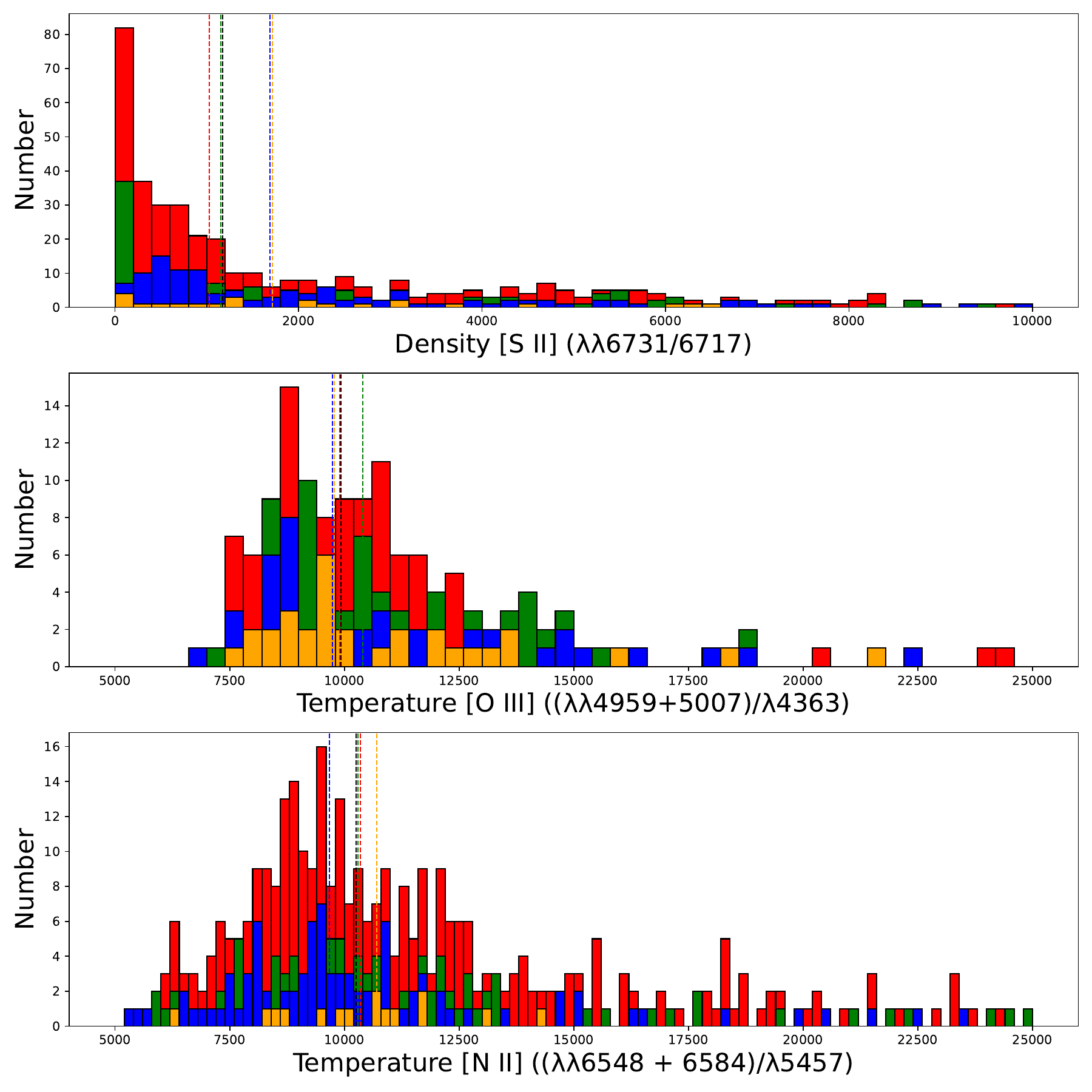}
    \caption{Histograms of {\it N$_e$} and {\it T$_e$} are plotted against the maximum number of sources obtained. The Galactic components are color-coded as follows: bulge (blue), thin disk (red), thick disk (green), and halo (orange). The dashed lines indicate the median values for each respective Galactic component, while the black line represents across all components.}
    \label{F:Te_Ne}
\end{figure}

The electron temperature, {\it T$_e$}, and electron density, {\it N$_e$}, for the ionized states were derived from the $H\alpha$ emission line ratios, initially assuming {\it T$_e$} = 10,000 K and {\it N$_e$} = 1,000 cm$^{-3}$ based on Case B conditions, as outlined by \cite{2006agna.book.....O}. Fig. \ref{F:Te_Ne} shows a histogram showing [S II] (${\lambda}$${\lambda}$6731/6717) {\it N$_e$} together with [O III] ((${\lambda}$${\lambda}$4959+5007)/${\lambda}$4363) and [N II] ((${\lambda}$${\lambda}$6548+6584)/${\lambda}$5457) {\it T$_e$}, which are emission lines in the optical region obtained with {\scshape neat} and expected in PNe. The c($H_\beta$), the {\it T$_e$} and {\it N$_e$} values of the PNe are given in Table \ref{T:Physic} (for all PNe, the table is given in Section \ref{sec:Supl}).

\subsection{Chemical Properties}

\begin{figure*}
\begin{center}
\includegraphics[width=\textwidth]{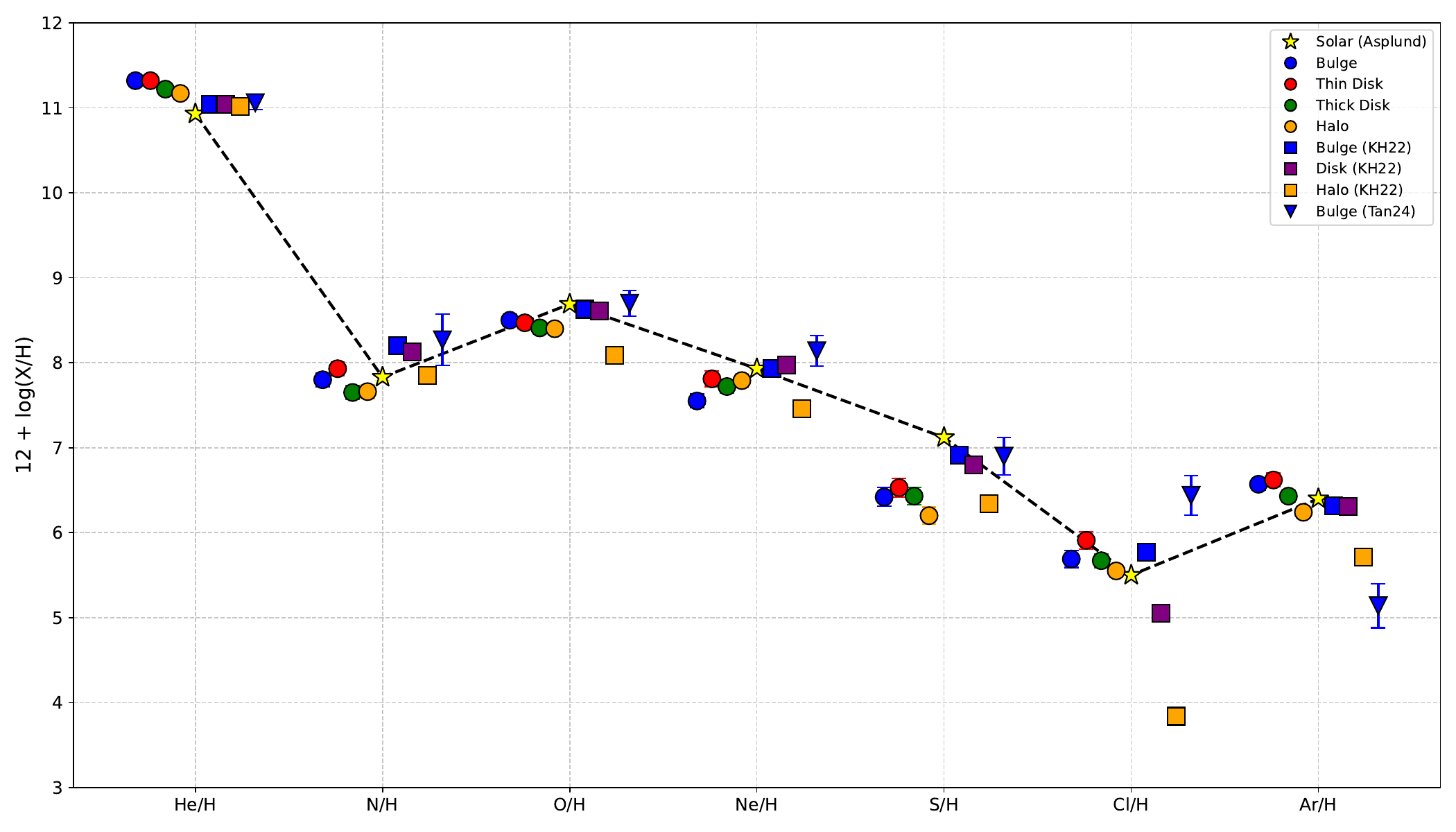}
\caption{The median elemental abundances of PNe presented in Table \ref{T:Abundance} are shown in the 12 + log(X/H) format for each element along the horizontal axis. Results from this study are represented by colored circles, while data from KH22 and Tan24 are indicated by square and inverted triangle symbols, respectively. The symbols are placed in arbitrary horizontal order for visual clarity. Solar abundance values are shown as yellow star symbols and connected by a dashed black line. Error bars are included for both this study and Tan24.}
\label{F:kwitter}
\end{center}
\end{figure*}

\begin{figure*}
\begin{center}
\includegraphics[width=\textwidth]{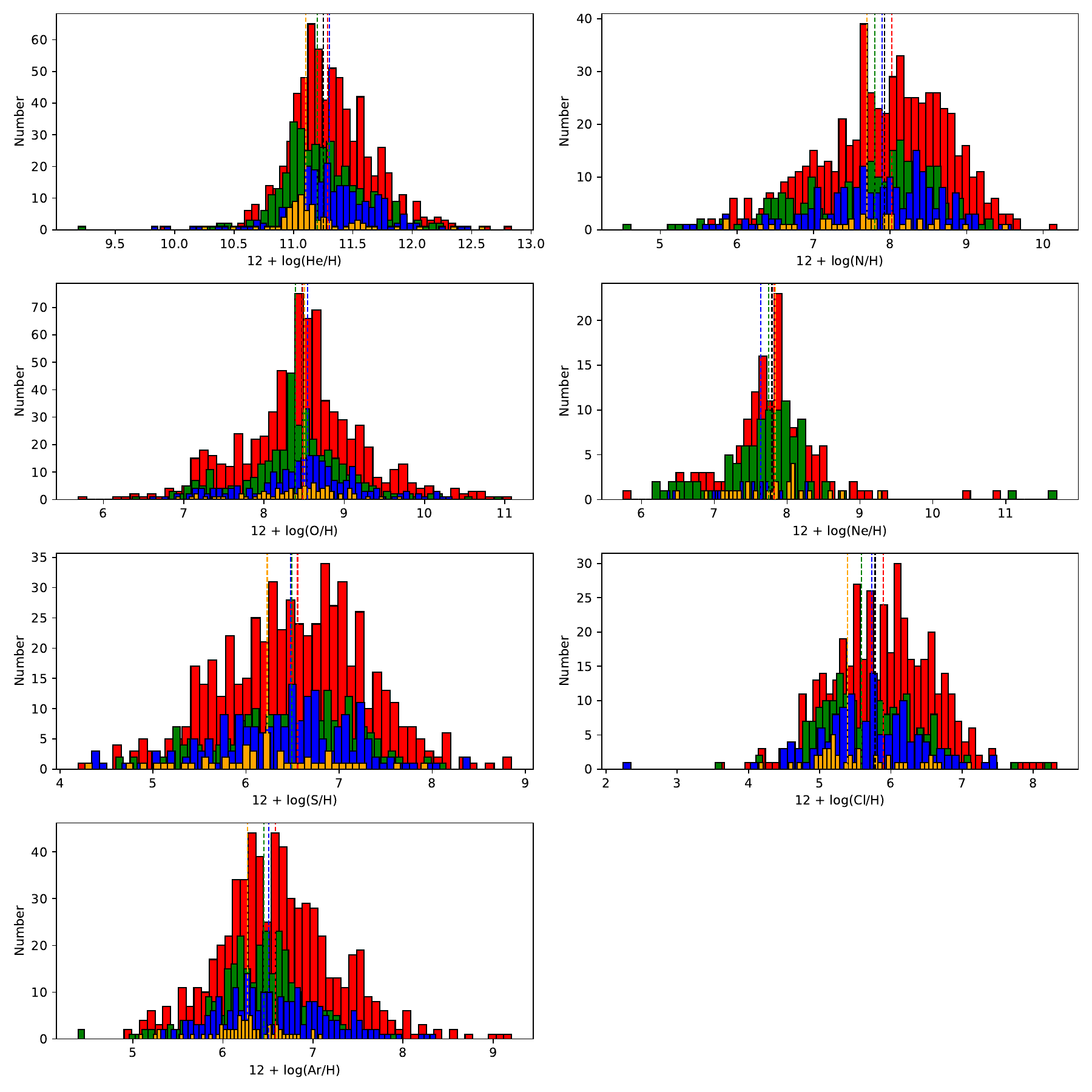}
\caption{Histograms of element abundances in log(X/H) + 12 form. The Galactic components are color-coded as follows: bulge (blue), thin disk (red), thick disk (green), and halo (orange). The dashed lines represent the median values of elemental abundances for respective Galactic components. The black line represents the median of the overall components.}
\label{F:elementabundancehistograms}
\end{center}
\end{figure*}

The abundance studies of PNe is crucial for understanding the interstellar medium and the evolutionary stages of the host star. Although many up-to-date programs have been developed, certain issues in elemental abundance calculations remain controversial, such as reddening corrections, the selection of appropriate densities and temperatures, corrections for unobservable ionic stages, and error analysis \citep{2023arXiv231201873S}. The ionic and elemental abundances of He, N, O, Ne, S, Cl, and Ar were derived using the {\scshape neat} code, based on emission line fluxes and their associated uncertainties. These uncertainties reflect only the statistical propagation of flux measurement errors and thus represent a lower bound on the uncertainties, as systematic effects were not taken into account. These element abundances are derived from collisionally excited lines (CELs) tuned to match the temperatures and densities associated with the ionization potential. When multiple lines are available for an ion, a weighted average is applied. For helium and heavier elements, recombination lines (RELs) are used to determine ionic abundances. \cite{2012MNRAS.425L..28P} provided the default logarithmic equation for He I abundances, which relates temperature and density and allows density calculations up to 1×10$^{14}$cm$^{-3}$ in the temperature range 5,000-25,000 K. Total elemental abundances were calculated by multiplying the ionic abundances of each element using the ICF scheme from \cite{2014MNRAS.440..536D}. 
It is important to note that ICFs, the underestimation of unobserved ion species, and the correlation effects between nebular dust and the central star can influence elemental abundances, potentially introducing significant uncertainties in abundance calculations \citep{2012ApJ...749...61H}. Table \ref{T:Abundance} is organized as follows: the first three raws present literature values of He, N, O, Ne, S, Cl and Ar abundances with respect to Galactic components from \cite{2009ARA&A..47..481A}(hereafter A19), Tan24 and KH22. The fourth row shows the average elemental abundances obtained in this study, while the remaining raws list some of the elemental abundances calculated for each PN (all are given in Section \ref{sec:Supl}). In addition, Fig. \ref{F:kwitter} average elemental abundances classified by the Galactic components and literature values using Table \ref{T:Abundance}. Moreover, In Fig. \ref{F:elementabundancehistograms} shows histograms of elemental abundances classified by the Galactic components.

\section{Results and Discussions}
\label{sec:res_dis}
We have carried out a comprehensive study of the physical and chemical properties of the PNe in the HASH database. The spectral datasets of 1,449 PNe located in the different components of the Galaxy were analyzed. 

The histograms in Fig. \ref{F:Size_Morpho} reveal a similar rapid decreasing trend in the angular sizes (in arcseconds and pc) of PNe in different Galactic components. The overall distributions have median angular sizes of 12 arcseconds and 0.30 parsecs. In addition, the angular sizes of the PNe in the halo are larger than those of the other components. However, as shown in the ERBIAS and sparm (Fig. \ref{F:Size_Morpho}c and Fig. \ref{F:Size_Morpho}d) classifications, the number of PNe in the thin disk is higher than in the other components, as expected. These histograms also indicate that the elliptical, bipolar, and asymmetric morphologies of PNe are the most predominant.

As shown in Fig. \ref{F:r_dist}, the distribution of distances between the central coordinates of the PNe and the nearest Gaia sources, categorized by Galactic components, reveals the following: 62\% of the sources are located within the 0–1 arcsec range, with the Thin Disk component (416 sources) being the most prevalent in this interval. Additionally, 87.5\% of the Halo component sources fall within this range. Beyond 1 arcsec, the number of sources drops sharply, exhibiting a right-skewed distribution.

The distances from the Gaia EDR3 exhibit a distribution of PNe given in Fig. \ref{F:dist} shows a right-skewed distribution with a median distance of 5 kpc and a maximum distance of 17 kpc the skewed part of this distribution contains mostly in the bulge. Parallax values generally range from 0.08 to 1.82 mas, while uncertainties range from 0.08 to 0.3 mas. The median parallax and uncertainty values are 0.21 mas and 0.20 mas, respectively. These millisecond-level uncertainties ensure that relative errors remain low, thereby reinforcing the reliability of our source selection. The parallax--distance relationship, which is expected to be inversely proportional to distance, is consistent with the findings of our study. As shown in Figure~\ref{F:r_parallax}, parallax values are measured well above 1 mas at closer distances (e.g., below 1 kpc), but decrease significantly with increasing distance; reaching approximately 0.16 mas at distances of 5--6 kpc and approximately 0.08 mas at distances of 6--7 kpc. This indicates that parallax measurements for distant sources become increasingly uncertain and, in some cases, drop to negative values. The Galactic distribution of PNe given in Fig. \ref{F:Aitoff_sky} is predominantly found in the disk (both thin and thick) and exhibits a similar distribution as shown in Fig. \ref{F:dist}.

The results of analysis given in Table \ref{T:DFlux} cover 31 distinct emission lines used to calculate the physical and chemical properties of the PNe. A total of 16,545 emission line measurements were obtained and presented in  Section \ref{sec:Supl}. While the H$\alpha$ flux is present in all PNe, only 4 PNe show the O II $\lambda$4089.29 line. The [O III] $\lambda$5006.84 line, the brightest line in PNe, is observed in 1,387 PNe, whereas the  [O III] $\lambda$4958.91 line is detected in 1,371 PNe. PN G078.3-02.7 and PN G356.1-03.3  have the highest number of emission lines, with 26 each, while no PN shows all the lines. In contrast, PN G000.3-03.0 and PN G359.2+04.7 stand out with the least number of lines, showing only $H_\alpha$ $\lambda$6562.77, [N II] $\lambda$6583.5, and [S II] $\lambda$6730.81. For flux measurements, only lines with signal-to-noise ratios (S/N) exceeding the 5-sigma threshold were included. The median S/N was found to be 9.4, while the median relative flux error (fluxerror / flux) was calculated to be 11\%. These fluxes and their associated errors were propagated in {\scshape neat} to derive elemental abundances and physical conditions. The resulting abundance uncertainties are approximately 10\%, as presented in Table \ref{T:Abundance}.

Plotting a large sample of PNe on the SMB and BPT diagnostic diagrams (see Fig. \ref{F:smb_bpt}) shows that the majority of them lie within the PN boundaries. However, some PNe are also found in regions associated with H II and SNRs, with only rare occurrences outside the PN boundaries. The SMB diagram shows a lower $H_\alpha$/[N II] ratio for PNe than H II regions and SNRs, while the BPT diagrams highlight a higher [O III] ($\lambda$5007)/H$\beta$  ratio, helping in their distinction \citep{2010PASA...27..129F, 2022FrASS...9.5287P}.

In Table \ref{T:Physic}, the c($H_\beta$) values are listed for 1,449 PNe. As shown in Fig. \ref{f:cbeta}, c($H_\beta$), representing the interstellar medium extinction calculated from the $H_\alpha$ / $H_\beta$ line ratios, exhibits a right-skewed distribution ranging from 0 to 6, with a median of 1.5. The median c($H_\beta$) values for the bulge and thin disk regions are higher than the overall median for all PNe. In addition, the median values of c($H_\beta$) for Galactic components are as follows: bulge (1.7), thin disk (1.5), thick disk (0.9), and halo (0.4).

For a total of 1,449 PNe, 891 {\it T$_e$} values and 889 {\it N$_e$} values are calculated using various emission lines, excluding cases with high temperatures (T$_e$ > 35,000 K) and densities ({\it N$_e$} < 1 cm$^{-3}$ and {\it N$_e$} > 10$^6$ cm$^{-3}$ ). Among these, 217 high-temperature values are identified, with 65 (30\%) attributed to [Ar III].
As [O III](${\lambda}$${\lambda}$4959+5007)/${\lambda}$4363, [N II] (${\lambda}$${\lambda}$6548 + 6584)/${\lambda}$5457), and [S II] ${\lambda}$${\lambda}$6731/6717) emission lines were measured for a large number of PNe, the histograms of {\it T$_e$} and {\it N$_e$} derived from these lines are presented in Fig. \ref{F:Te_Ne}.
The histogram of {\it N$_e$} shows a rapidly decreasing trend from 1 to 1,000 cm$^{-3}$ with a median value of 1,170 cm$^3$. 
The higher to lower density for Galactic components are halo (1,715 cm$^{-3}$) to thin disk (1,030 cm$^{-3}$). 
The {\it T$_e$} distribution shows similar Gaussian patterns with median values of 9,920 K and 10,250 K, aligning with the typical temperature of 10,000 K for a PN as given by \cite{2006agna.book.....O}. We can interpret that the temperature and density distributions do not vary significantly across Galactic components. However, the distributions of temperature and density are distinct from each other.

The abundance values determined in this study are primarily shown in Fig. \ref{F:kwitter} for comparison with KH22 (Figs. 5–7), Tan24 and A19 for solar values. This figure presents the average abundances (from Table \ref{T:Abundance}) of He, N, O, Ne, S, Cl, and Ar (Hereafter, the format of element abundances is used as X instead of X/H) in PNe. The element abundance distributions across the Galactic components follow a similar trend. Notably, Tan24, representing bulge values, shows higher element abundances, except for Ar. In contrast, the halo values for Cl and Ar in KH22 are significantly lower. Furthermore, the S abundances identified in this study are lower than the solar value. The uncertainties in the abundance values for both this study and Tan24 are indicated by error bars in Fig. \ref{F:kwitter}.

The elemental abundance distributions of PNe for He, N, O, Ne, S, Cl, and Ar display Gaussian-like shapes (Fig. \ref{F:elementabundancehistograms}), with median values of 11.3, 8.0, 8.4, 8.6, 6.6, 5.8, and 6.5, respectively. PNe in the thin disc exhibit high abundances above the median of the Galactic components (except O and Ne). On the other hand, the abundances of PNe in the halo and thick disc (except O) are below the median. Among the elements given above, He has the highest abundance, and Cl the lowest. In the bulge, abundance values align closely with the median, though He and O are relatively higher and Ne is lower by 0.15 dex. The explanations of the plots based on the median values are given below:\\
\textbf{He:} It is lower in the halo and higher in the bulge.\\
\textbf{N:} The value is higher in the thin disk and lower in the halo.\\
\textbf{O:} It is consistent across components, with a slightly lower value in the thick disk.\\
\textbf{Ne:} It has fewer data points, but the mean values across components are similar.\\
\textbf{S:} The halo value is approximately 0.3 dex lower than in other components.\\
\textbf{Cl:} It is relatively higher in the thin disk.\\
\textbf{Ar:} It is lower in the halo and higher in the thick disk.\\

\begin{figure*}
\begin{center}
\includegraphics[width=\textwidth]{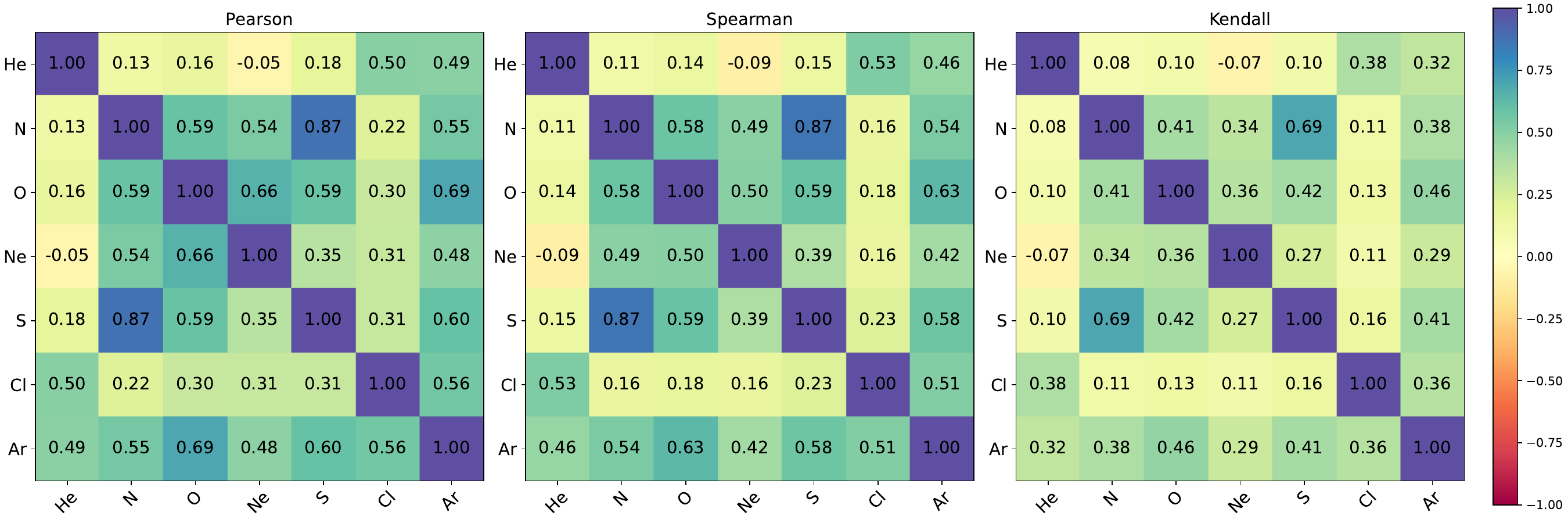}
\caption{Pearson (r), Spearman ($\rho$), and Kendall ($\tau$) correlation coefficients are calculated for the elemental abundances (ratio to H) for PNe. The color index is described as follows: Blue indicates a positive correlation, ranging from 0.00 to a maximum of 1.00, while red represents a negative correlation, ranging from 0.00 to a minimum of -1.00.}
\label{F:corr}
\end{center}
\end{figure*}

\begin{figure*}
\begin{center}
\includegraphics[width=\textwidth]{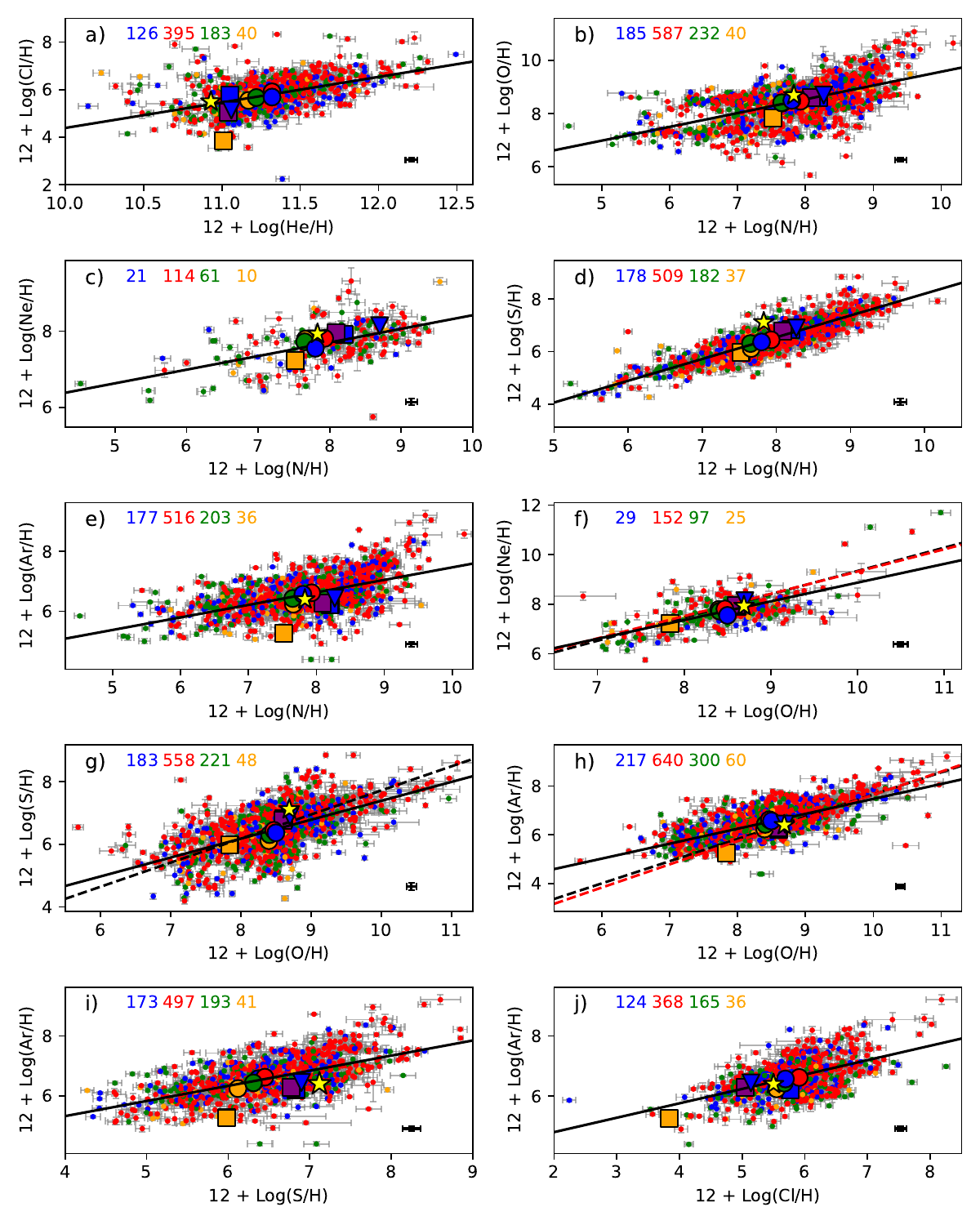}
\caption{Elemental abundances of PNe with correlation coefficients ranging from 0.5 to 1. Values are on a scale of 12 + log(X/H). The blue, red, green, and orange circles represent the mean elemental abundances of the bulge region, thin disk, thick disk, and halo regions, respectively. The blue, purple, and orange squares represent the abundance ratios of the bulge, disk and halo, respectively from KH22 and the yellow star symbol represents the solar abundance in A19. The blue triangle indicates the abundance ratio for the bulge region in the Tan24 study. The solid and dashed lines represent the least-squares fit of the data of our study and KH22, respectively, with the properties provided in Table \ref{T:Abundance_R}. Error bars, representing mean uncertainties, are displayed in the lower-right corner of each panel, while individual error bars are included for each abundance measurement.}
\label{F:Abundance_plot}
\end{center}
\end{figure*}

Elemental abundances of 1,449 PNe were analyzed using Pearson ($r$, \citealp{pearson1895vii}), Spearman ($\rho$, \citealp{spearman1961proof}) and Kendall ($\tau$, \citealp{kendall1975}) correlation tests to evaluate their interrelationships, with the complete statistical analysis presented in Fig. \ref{F:corr}. Since the three statistical methods (Pearson, Spearman and Kendall) produced very similar coefficient values and Pearson's correlation is widely used in the literature, we selected Pearson's statistics for our study. The color-coded heatmap highlights positive correlations in blue, with darker shades indicating stronger correlation, while red areas (representing negative correlations) are minimal. Pearson's correlation (r) measures linear correlation: r > 0.5 indicates strong positive, 0.3–0.5 moderate positive, and 0.0–0.3 weak positive correlations, while negative values show inverse correlation. As shown in Fig. \ref{F:corr}, a total of 21 correlation coefficients were found out of the diagonal correlation coefficients, comprising 10 strong, 6 moderate, 4 weak, and 1 negative weak correlation ($r = -0.05$, Ne/He). It is noted that Ar is strongly correlated with 4 out of 6 elements. Also, the strongest positive correlation is found for S/N ($r = 0.87$). For correlations $r > 0.3$, a least-squares fit was applied, with equations and parameters detailed in Table \ref{T:Abundance_R}. The abundance correlation plots are provided in Fig. \ref{F:Abundance_plot} for $r > 0.5$ and Fig. \ref{F:A1} for $0.3> r > 0.5$. The abundance comparisons expressed as 12 + log(X/H) offer insights into the interstellar medium, incorporating average values from A19, KH22, and Tan24.

Table \ref{T:Abundance_R} presents the least-squares fit parameters, Pearson correlation coefficient, and the number of data points used in the fits for 16 cross correlations across each Galactic component. All slope values are positive. The highest and lowest correlation coefficient values among the Galactic components were observed for the S/N and O/Cl abundance ratios, respectively. The O/Ar ratio was analyzed using the largest sample of PNe (1,217), whereas the N/Ne ratio had the smallest sample size (206).

Strong correlations (r > 0.5) for Cl/He, O/H, Ne/N, S/N, Ar/N, Ne/O, S/O, Ar/O, Ar/S, and Ar/Cl were used to construct Fig. \ref{F:Abundance_plot}. However, the scatter in these distributions is attributed to uncertainties in measured line intensities, reddening corrections, and the absence of certain ionic lines. The mean values of these elements for the Galactic components are presented in these plots for comparison, along with the solar value and the values reported by KH22 and Tan24. In these plots, the most notable data is the slight scatter of the halo values reported in KH22. On the other hand, for Ar/Cl, the scattering in the KH22 data for the halo is about 2 dex. We attribute this difference to the fact that KH22's sample includes a mean abundance of only 12 Ar and 4 Cl out of 13 PNe.

For the Ne/N plot, the value from Tan24 for the bulge was observed to be 1 dex. larger than our bulge value and 0.5 dex. larger than the bulge value from KH22.

The strongest positive correlation belongs to S/N ($r = 0.87$), which stands out significantly from the other distributions.
A slope of 0.83 in the S/H versus N/H plot indicates that sulfur increases with nitrogen but at a slower rate. This reflects their distinct origins: nitrogen is enriched through the CNO cycle and s-process nucleosynthesis in intermediate- and low-mass stars during the AGB phase, particularly in massive stars that evolve into Type I PNe. Sulfur, an alpha-element, is largely unaffected by these processes, reflecting the interstellar medium's composition at the progenitor star's formation. This slower increase in S/H likely highlights sulfur's independence from stellar nucleosynthesis, while nitrogen is more strongly tied to stellar evolution, potentially allowing differentiation of PNe by progenitor mass and evolution.

Linear fit values for the Ne/O, S/O, and Ar/O ratios from the KH22 were compared with those derived for the Galactic components analyzed in this study. A linear fit with a correlation coefficient of r = 0.7 between Ne/O is observed, despite limited contributions from components other than the thin disk. Considering all Galactic components, the slope of the linear fit in KH22 is 0.95 (r = 0.72), whereas we determined it to be 0.76 (r = 0.66). It should be noted that we obtained this correlation with 3 times more data. 

The S/O distribution exhibits a slope of a linear fit of 0.60 (r = 0.59) comparable to that of 0.77 (r = 0.47) KH22. The mean values for the Galactic components are similar to each other but are 0.5 dex. lower than the solar values along both axes.

For the Ar/O distribution, the slope of the linear equation obtained from KH22 is 0.91 (R = 0.66), whereas the slope of the equation covering all components is 0.61 (R = 0.69). A similar discrepancy, as observed in the Ne/O distribution, can be attributed to the sample size difference, which is a factor of 10. The average Ar/H values of the Galactic components are close to the Sun, KH22 and Tan24. The most notable observation in this graph is that the Ar/O values of PNe in the halo differ by 1 dex. from both the Solar and Galactic values.

Fig. \ref{F:A1} shows the elemental abundance plots with moderate positive (0.3 > r > 0.5) correlations for Ar/He, Cl/O, S/Ne, Cl/Ne, Ar/Ne, and Cl/S. Among these, the Ar/He plot has the largest number of PNe (1,168). Similarly in Figure \ref{F:Abundance_plot}, halo values for different elements in KH22 are scattered in all plots, while other component values are concentrated around 0.5 dex. 
Among these plots, the Cl/O distribution is the only one that shows a significant difference in slope compared to KH22. In KH22, the Cl/O distribution showed a slope of 0.83 (R = 0.56), while in this study, it exhibited a slope of 0.31 (R = 0.30). Notably, our dataset is seven times larger in this correlation.

\section{Conclusions}
\label{sec:con}

A comprehensive analysis of 1,449 out of 2,591 PNe from the HASH database was conducted in this study to examine their physical and chemical properties across different Galactic components. This study revealed that PNe angular sizes and morphologies, including elliptical, bipolar, and asymmetric types, were predominantly distributed within the Galactic disk, with median angular sizes of 12 arcseconds and 0.45 parsecs. Also, the Halo PNe exhibit larger angular dimensions, indicating distinct characteristics compared to other components.

The physical parameters c($H_\beta$), T$_e$, and N$_e$, derived from most of the 31 emission lines across all PNe, exhibit Gaussian-like distributions with median values of 1.5, 9,900 K, and 1,200 cm$^{-3}$, respectively. The distributions of these parameters across Galactic components showed slight variations. The c($H_\beta$) values in the bulge and thin disk regions are higher than the overall median for PNe.

The elemental abundance distributions for H (12.0), He (11.3), N (7.8), O (8.4), Ne (7.7), S (6.3), Cl (5.7) and Ar (6.5) show Gaussian-like patterns, here the mean values are given in parentheses. Thin disk PNe show higher abundances than average, except for O and Ne, while halo PNe have the lowest values. The bulge aligns closely with Galactic averages but exhibits slightly elevated He and O and lower Ne.

Statistical analyses, including Pearson, Spearman, and Kendall correlation tests, revealed strong correlations between elements, with Sulfur and Nitrogen showing a particularly high correlation (r = 0.87), accounting for half of the total 21 correlations. Comparison with KH22 and Tan24 reveals discrepancies (< 2 dex.) in elemental abundance ratios, particularly in the halo.

As noted, the bias uncertainties were approximately 15.5\%, impacting only a small fraction of the dataset. Additionally, the flux measurements carry uncertainties of about 11\%. Overall, these biases and uncertainties are unlikely to have a significant effect on the results.

This study underscores the importance of leveraging large datasets to uncover trends in PNe properties and provides a foundation for future investigations into their role in Galactic chemical evolution. In the near future, we aim to expand the study to include Large and Small Magellanic Clouds. Additionally, we plan to utilize the findings from this study, along with photoionization models, to gain insights into progenitor stars.

\section*{Acknowledgements}

This research was supported by the Scientific and Technological Research Council of Turkey (TÜBİTAK) through project number 122F122. We thank to Sinan Kaan Yerli to help downloading data. This work has made use of data from the European Space Agency (ESA) mission Gaia (www.cosmos.esa.int/gaia), processed by the Gaia Data Processing and Analysis Consortium (DPAC, www.cosmos.esa.int/web/gaia/dpac/consortium). Funding for the DPAC has been provided by national institutions, in particular, the institutions participating in the Gaia Multilateral Agreement. During the preparation of this work, the author(s) utilized ChatGPT-4o to refine the English wording of certain phrases. Following the use of this tool, the author(s) reviewed and edited the content as necessary and took full responsibility for the content of the publication.

\section*{Data Availability}

The data may be available from the corresponding author on request.

\section*{Supplementary Material}
\label{sec:Supl}

Additional supporting information, including four files, is available in the online version of this article and within it.

\textbf{\texttt{Table1-Full.csv}}:
The Complete version of Table \ref{T:HASH}, properties for all PNe.

\textbf{\texttt{Table2-Full.csv}}:
The complete version of Table \ref{T:DFlux}, reddening corrected emission line fluxes for all PNe.

\textbf{\texttt{Table3-Full.csv}}:
The complete version of Table \ref{T:Physic}, The c(H$_{\beta}$), {\it T$_e$} and {\it N$_e$} values of all PNe.

\textbf{\texttt{Table4-Full.csv}}:
The complete version of Table \ref{T:Abundance}, The elemental abundances of all PNe.



\bibliographystyle{mnras}
\bibliography{pne} 



\begin{table*}
\centering
\caption{Investigated True PNe sample from the HASH database. The columns are as follows; HASH ID, PNG, Name, DRA, DDEC, Glon, Glat, MajDiam, MinDiam, mainClass, and subClass information. The table is sorted by HASH ID. \textit{Rgeo} data are taken from Gaia Data \protect\citep{2021AJ....161..147B}. In this and the next three tables, only 13 sources are provided. The complete table can be found in Section \ref{sec:Supl}.}
    \begin{tabular}
     {cccccccccccc}
    \hline\hline
    HASH ID	&	PN G	&	Name	&	DRA	&	DDEC	&	\textit{l} 	&	\textit{b}	&	MajDiam	&	MinDiam	&	mainClass	&	subClass  &	 \textit{Rgeo}	\\
    &&&    (J2000) &   (J2000)    &    (deg) & (deg) &&&&&(kpc)  \\
\hline
11	&	000.1-01.1	&	M 3-43	&	267.6	&	-29.42	&	0.12	&	-1.15	&	3.5	&	2.7	&	E	&	-	&	6.84	\\
13	&	000.1-05.6	&	H 2-40	&	272.13	&	-31.61	&	0.15	&	-5.62	&	18.3	&	16.9	&	E	&	a	&	5.66	\\
14	&	000.1+17.2	&	PC 12	&	250.97	&	-18.95	&	0.17	&	17.25	&	2.3	&	2.2	&	B	&	m	&	7.82	\\
17	&	000.1-02.3	&	Bl 3-10	&	268.84	&	-29.96	&	0.2	&	-2.34	&	7.2	&	6.9	&	R	&	a	&	6.7	\\
19	&	000.2-01.9	&	M 2-19	&	268.44	&	-29.73	&	0.23	&	-1.93	&	20	&	-	&	B	&	-	&	9.03	\\
20	&	000.2-04.6	&	Wray 16-363	&	271.18	&	-31.05	&	0.26	&	-4.64	&	6.4	&	-	&	E	&	m	&	6.85	\\
22	&	000.3-04.6	&	M 2-28	&	271.26	&	-30.97	&	0.36	&	-4.66	&	5.5	&	-	&	E	&	-	&	6.8	\\
23	&	000.3+12.2	&	IC 4634	&	255.39	&	-21.83	&	0.36	&	12.21	&	20.5	&	6.6	&	B	&	m	&	2.46	\\
25	&	000.3-02.8	&	M 3-47	&	269.43	&	-30.04	&	0.39	&	-2.83	&	9	&	8	&	R	&	-	&	8.67	\\
26	&	000.4-01.9	&	M 2-20	&	268.61	&	-29.60	&	0.41	&	-1.99	&	6	&	-	&	B	&	-	&	10.47	\\
27	&	000.4-02.9	&	M 3-19	&	269.58	&	-30.01	&	0.48	&	-2.93	&	7.2	&	6.6	&	R	&	-	&	7.63	\\
29	&	000.5+01.9	&	JaSt 17	&	264.88	&	-27.46	&	0.55	&	1.92	&	9.1	&	6.6	&	E	&	-	&	1.46	\\
30	&	000.5-03.1	&	KFL 1	&	269.81	&	-30.05	&	0.55	&	-3.12	&	8	&	7.9	&	R	&	-	&	7.69	\\
\dots & \dots & \dots & \dots & \dots & \dots & \dots & \dots & \dots & \dots & \dots & \dots \\
    \hline
    \end{tabular}
\label{T:HASH}
\end{table*}

\begin{table*}
\centering
\caption{The reddening corrected emission line fluxes of PNe. All line fluxes are scaled to H$\beta$ fluxes (= 100). "-": Refers to emission lines that could not be detected.}
    \begin{tabular}{ccc@{~~}c@{~~}c@{~~}c@{~~}c@{~~}c@{~~}c@{~~}c@{~~}c@{~~}c@{~~}c@{~~}c@{~~}c@{~~}c@{~~}c@{~~}c@{~~}c@{~~}c@{~~}c@{~~}c@{~~}c@{~~}c@{~~}c@{~~}c@{~~}c@{~~}c@{~~}c@{~~}c@{~~}c@{~~}c@{~~}c@{~~}}
    \hline\hline
    HASH ID & PN G	&	$[\text{O II}]$	&	$[\text{O II}]$	&	$\text{O II}$	&	H$_{\delta}$	&	H$_{\gamma}$	&	$[\text{O III}]$	&	$\text{He I}$	&	$\text{O II}$	&	$\text{O II}$	&	$[\text{Ar IV}]$	&	$[\text{Ar IV}]$	&	$[\text{O III}]$	&	$[\text{O III}]$	&	$[\text{Ar III}]$	&	$[\text{Cl III}]$	&	$[\text{Cl III}]$	&	$[\text{O I}]$	&	$[\text{N II}]$	&	$\text{He I}$	&	$[\text{O I}]$	&	$[\text{O I}]$	&	$[\text{N II}]$	&	H$_{\alpha}$	&	$[\text{N II}]$	&	$\text{He I}$	&	$[\text{S II}]$	&	$[\text{S II}]$	&	$[\text{Ar V}]$	&	$[\text{Ar III}]$	&	$\text{He I}$	&	$[\text{Ar III}]$	\\
 &  &	$\lambda$3726.03	&	$\lambda$3728.82	&	$\lambda$4089.29	&	$\lambda$4101.74	&	$\lambda$4340.47	&	$\lambda$4363.21	&	$\lambda$4471.50	&	$\lambda$4649.13	&	$\lambda$4661.63	&	$\lambda$4711.37	&	$\lambda$4740.17	&	$\lambda$4958.91	&	$\lambda$5006.84	&	$\lambda$5191.82	&	$\lambda$5517.66	&	$\lambda$5537.60	&	$\lambda$5577.34	&	$\lambda$5754.60	&	$\lambda$5875.66	&	$\lambda$6300.34	&	$\lambda$6363.78	&	$\lambda$6548.10	&	$\lambda$6562.77	&	$\lambda$6583.50	&	$\lambda$6678.16	&	$\lambda$6716.44	&	$\lambda$6730.82	&	$\lambda$7005.67	&	$\lambda$7135.80	&	$\lambda$7281.35	&	$\lambda$7751.06	\\
     \hline
11	&	000.1-01.1	&	-	&	-	&	-	&	-	&	-	&	-	&	-	&	-	&	-	&	-	&	-	&	373	$\pm$	60	&	1,130	$\pm$	190	&	-	&	-	&	2.78	$\pm$	0.48	&	-	&	16.80	$\pm$	2.00	&	18.40	$\pm$	2.70	&	7.01	$\pm$	1.13	&	153	$\pm$	25	&	277	$\pm$	42	&	393	$\pm$	65	&	6	$\pm$	1.01	&	18.20	$\pm$	3.25	&	32.40	$\pm$	5.60	&	3.41	$\pm$	0.65	&	41.00	$\pm$	8.25	&	-	&	-	\\
13	&	000.1-05.6	&	-	&	-	&	-	&	-	&	-	&	-	&	-	&	-	&	-	&	-	&	-	&	140	$\pm$	13	&	438	$\pm$	23	&	-	&	-	&	-	&	-	&	-	&	-	&	-	&	115	$\pm$	22	&	285	$\pm$	38	&	384	$\pm$	61	&	-	&	50.10	$\pm$	10.00	&	45.90	$\pm$	8.95	&	-	&	32.50	$\pm$	7.15	&	30.80	$\pm$	7.70	&	-	\\
14	&	000.1+17.2	&	-	&	-	&	-	&	-	&	26.50	$\pm$	0.10	&	-	&	4.07	$\pm$	0.41	&	-	&	-	&	-	&	-	&	110	$\pm$	14	&	386	$\pm$	49	&	-	&	-	&	-	&	-	&	26.30	$\pm$	2.30	&	8.63	$\pm$	0.97	&	3.19	$\pm$	0.44	&	-	&	1,100	$\pm$	165	&	381	$\pm$	50	&	10	$\pm$	1.73	&	-	&	25.80	$\pm$	3.10	&	-	&	47.60	$\pm$	6.30	&	-	&	-	\\
17	&	000.1-02.3	&	-	&	-	&	-	&	25.50	$\pm$	2.60	&	-	&	-	&	-	&	-	&	-	&	-	&	-	&	524	$\pm$	39	&	1,660	$\pm$	90	&	-	&	-	&	12.90	$\pm$	2.30	&	-	&	14.00	$\pm$	1.70	&	-	&	-	&	-	&	829	$\pm$	59	&	41	$\pm$	5	&	-	&	-	&	12.90	$\pm$	2.60	&	29.10	$\pm$	3.90	&	77.90	$\pm$	8.50	&	-	&	-	\\
19	&	000.2-01.9	&	-	&	-	&	-	&	-	&	-	&	-	&	-	&	-	&	-	&	-	&	-	&	72	$\pm$	6	&	259	$\pm$	13	&	-	&	-	&	-	&	-	&	48.90	$\pm$	2.30	&	-	&	-	&	-	&	1,260	$\pm$	105	&	624	$\pm$	48	&	14	$\pm$	2.20	&	44.70	$\pm$	6.80	&	70.50	$\pm$	6.10	&	-	&	48.60	$\pm$	5.05	&	-	&	-	\\
20	&	000.2-04.6	&	-	&	-	&	-	&	-	&	21.00	$\pm$	0.10	&	-	&	-	&	-	&	-	&	-	&	-	&	224	$\pm$	33	&	706	$\pm$	99	&	-	&	-	&	-	&	66.00	$\pm$	4.00	&	41.60	$\pm$	7.35	&	65.80	$\pm$	4.00	&	20.10	$\pm$	3.15	&	328	$\pm$	12	&	752	$\pm$	37	&	978	$\pm$	53	&	24	$\pm$	1.65	&	80.40	$\pm$	9.10	&	81.60	$\pm$	11.60	&	-	&	-	&	-	&	-	\\
22	&	000.3-04.6	&	-	&	-	&	-	&	6.47	$\pm$	0.51	&	21.10	$\pm$	0.10	&	1.72	$\pm$	0.27	&	5.25	$\pm$	0.59	&	-	&	-	&	-	&	-	&	221	$\pm$	25	&	690	$\pm$	71	&	-	&	1.95	$\pm$	0.30	&	-	&	-	&	19.39	$\pm$	0.66	&	22.10	$\pm$	2.00	&	-	&	186	$\pm$	4	&	298	$\pm$	6	&	578	$\pm$	9	&	5	$\pm$	0.33	&	35.50	$\pm$	1.30	&	45.90	$\pm$	1.70	&	2.06	$\pm$	0.38	&	30.70	$\pm$	1.30	&	-	&	-	\\
23	&	000.3+12.2	&	-	&	-	&	-	&	19.30	$\pm$	1.70	&	41.80	$\pm$	0.10	&	4.40	$\pm$	0.51	&	3.43	$\pm$	0.34	&	-	&	-	&	1.97	$\pm$	0.01	&	-	&	348	$\pm$	26	&	1,110	$\pm$	60	&	-	&	1.00	$\pm$	0.19	&	-	&	-	&	19.80	$\pm$	1.20	&	-	&	-	&	-	&	425	$\pm$	31	&	15	$\pm$	1	&	5	$\pm$	0.42	&	-	&	-	&	-	&	15.90	$\pm$	1.25	&	1.89	$\pm$	0.28	&	-	\\
25	&	000.3-02.8	&	-	&	-	&	-	&	-	&	125.00	$\pm$	1.00	&	-	&	-	&	-	&	-	&	-	&	-	&	216	$\pm$	22	&	771	$\pm$	21	&	-	&	-	&	-	&	-	&	77.60	$\pm$	8.90	&	68.20	$\pm$	9.10	&	-	&	689	$\pm$	70	&	1,220	$\pm$	80	&	2,210	$\pm$	100	&	-	&	207.00	$\pm$	14.00	&	184.00	$\pm$	16.00	&	-	&	162.00	$\pm$	11.00	&	-	&	-	\\
26	&	000.4-01.9	&	-	&	-	&	-	&	-	&	23.10	$\pm$	0.10	&	-	&	-	&	-	&	5.79	$\pm$	1.13	&	6.79	$\pm$	0.01	&	-	&	215	$\pm$	16	&	697	$\pm$	38	&	4.57	$\pm$	0.64	&	3.07	$\pm$	0.35	&	-	&	4.62	$\pm$	0.92	&	61.60	$\pm$	3.50	&	10.90	$\pm$	1.30	&	-	&	-	&	1,250	$\pm$	90	&	466	$\pm$	29	&	22	$\pm$	1.90	&	20.60	$\pm$	2.70	&	36.30	$\pm$	2.70	&	-	&	119.00	$\pm$	8.00	&	5.26	$\pm$	0.96	&	-	\\
27	&	000.4-02.9	&	-	&	-	&	-	&	-	&	31.60	$\pm$	0.10	&	-	&	-	&	-	&	-	&	-	&	-	&	165	$\pm$	14	&	608	$\pm$	10	&	-	&	-	&	-	&	-	&	38.00	$\pm$	3.55	&	-	&	-	&	-	&	649	$\pm$	73	&	63	$\pm$	11	&	11	$\pm$	1.15	&	-	&	7.20	$\pm$	1.10	&	-	&	31.40	$\pm$	4.40	&	-	&	-	\\
29	&	000.5+01.9	&	-	&	-	&	-	&	-	&	27.80	$\pm$	0.10	&	-	&	-	&	-	&	-	&	-	&	-	&	189	$\pm$	9	&	672	$\pm$	23	&	-	&	-	&	19.00	$\pm$	3.20	&	-	&	144.00	$\pm$	6.00	&	-	&	-	&	-	&	3,080	$\pm$	140	&	-	&	77	$\pm$	3.60	&	-	&	-	&	16.60	$\pm$	1.55	&	239.00	$\pm$	10.00	&	-	&	91.70	$\pm$	0.10	\\
30	&	000.5-03.1	&	-	&	-	&	-	&	4.91	$\pm$	0.57	&	21.00	$\pm$	0.10	&	2.69	$\pm$	0.53	&	4.28	$\pm$	0.55	&	-	&	-	&	-	&	-	&	240	$\pm$	38	&	769	$\pm$	120	&	-	&	1.79	$\pm$	0.33	&	-	&	-	&	38.29	$\pm$	0.32	&	-	&	-	&	-	&	845	$\pm$	7	&	40	$\pm$	2	&	17	$\pm$	0.66	&	8.58	$\pm$	0.46	&	8.31	$\pm$	0.59	&	1.43	$\pm$	0.25	&	43.37	$\pm$	0.97	&	-	&	-	\\
\dots & \dots & \dots & \dots & \dots & \dots & \dots & \dots & \dots & \dots & \dots & \dots & \dots \\
    \hline
    
    \hline
    \end{tabular}
    \label{T:DFlux}
\end{table*}

\begin{table*}
\centering
\caption{The c(H$_{\beta}$), {\it T$_e$} and {\it N$_e$} values of the PNe. They are sorted according to the HASH ID number. In the continuation of the table, the symbols ‘L’ and ‘H’ represent the lower and upper limits of temperature and density, respectively.}
    \begin{tabular}{c@{~~~}c@{~~}c@{~~}c@{~~}c@{~~}c@{~~}c@{~~}c@{~~}c@{~~}c@{~~}c@{~~}}
    \hline\hline
     \multicolumn{2}{c|}{}  & \multicolumn{5}{c|}{Electron Density [cm$^{-3}$]} & \multicolumn{4}{c@{~~}}{Electron Temperature [K]}\\
     \hline
     & 	&	c(H$_{\beta}$)	&	[O II] 	&	$[\text{S II}]$ 	&	$[\text{Cl III}]$ 	&	$[\text{Ar IV}]$  &	$[\text{O II}]$ 	&	$[\text{N II}]$ 	&	$[\text{O III}]$ 	&	$[\text{Ar III}]$	\\
     \hline
This Work	&	Bulge	&	1.77	$\pm$	0.13	&	545	$\pm$	209	&	3,412	$\pm$	1,076	&	3,939	$\pm$	1,772	&	5,210	$\pm$	1,817	&		-		&	11,512	$\pm$	1,155	&	10,987	$\pm$	514	&	18,462	$\pm$	2,438	\\
	&	Thin Disk	&	1.64	$\pm$	0.15	&	5,022	$\pm$	2,342	&	3,347	$\pm$	1,163	&	5,007	$\pm$	2,086	&	4,095	$\pm$	1,853	&	13,623	$\pm$	2,264	&	11,968	$\pm$	1,091	&	10,337	$\pm$	424	&	18,773	$\pm$	2,744	\\
	&	Thick Disk	&	0.94	$\pm$	0.13	&	2,713	$\pm$	977	&	3,311	$\pm$	1,144	&	3,914	$\pm$	1,780	&	336	$\pm$	168	&	18,944	$\pm$	4,684	&	11,723	$\pm$	1,064	&	11,019	$\pm$	417	&	15,618	$\pm$	1,920	\\
	&	Halo	&	0.44	$\pm$	0.12	&	1,300	$\pm$	537	&	2,735	$\pm$	1,069	&	4,183	$\pm$	1,685	&		-		&	14,850	$\pm$	2,350	&	12,632	$\pm$	1,314	&	11,004	$\pm$	431	&	22,650	$\pm$	2,625	\\
\hline
     HASH ID & PN G &&&&&&&&&\\
\hline
11	&	000.1-01.1	&	3.64	$\pm$	0.22	&	-	&	5,520	$\pm$	1,410	&	-	&	-	&	-	&	14,600	$\pm$	750	&	-	&	-	\\
13	&	000.1-05.6	&	0.69	$\pm$	0.2	&	-	&	448	$\pm$	364	&	-	&	-	&	-	&	-	&	-	&	-	\\
14	&	000.1+17.2	&	1.81	$\pm$	0.26	&	-	&	-	&	-	&	-	&	-	&	7,700	$\pm$	545	&	-	&	-	\\
17	&	000.1-02.3	&	1.44	$\pm$	0.14	&	-	&	-	&	-	&	-	&	-	&	-	&	-	&	-	\\
19	&	000.2-01.9	&	2.02	$\pm$	0.16	&	-	&	3,070	$\pm$	2,140	&	-	&	-	&	-	&	9,360	$\pm$	700	&	-	&	-	\\
20	&	000.2-04.6	&	1.32	$\pm$	0.24	&	-	&	707	$\pm$	511	&	-	&	-	&	-	&	-	&	-	&	-	\\
22	&	000.3-04.6	&	0.04	$\pm$	0.02	&	-	&	1,540	$\pm$	260	&	-	&	-	&	-	&	8,860	$\pm$	220	&	7,780	$\pm$	330	&	-	\\
23	&	000.3+12.2	&	0.53	$\pm$	0.14	&	-	&	-	&	-	&	-	&	-	&	22,300	$\pm$	3,200	&	8,780	$\pm$	290	&	-	\\
25	&	000.3-02.8	&	1.97	$\pm$	0.19	&	-	&	382	$\pm$	229	&	-	&	-	&	-	&	-	&	-	&	-	\\
26	&	000.4-01.9	&	2.08	$\pm$	0.13	&	-	&	4,280	$\pm$	865	&	-	&	-	&	-	&	8,270	$\pm$	450	&	-	&	24,800	$\pm$	3,550	\\
27	&	000.4-02.9	&	1.19	$\pm$	0.18	&	-	&	-	&	-	&	-	&	-	&	22,400	$\pm$	4,900	&	-	&	-	\\
29	&	000.5+01.9	&	3.23	$\pm$	0.1	&	-	&	-	&	-	&	-	&	-	&	-	&	-	&	-	\\
30	&	000.5-03.1	&	1.46	$\pm$	0.24	&	-	&	753	$\pm$	315	&	-	&	-	&	-	&	21,500	$\pm$	1,650	&	8,490	$\pm$	525	&	-	\\
\dots & \dots & \dots & \dots & \dots & \dots & \dots & \dots & \dots & \dots & \dots \\
    \hline
    
    \hline
    \end{tabular}
    \label{T:Physic}
\end{table*}

\begin{table*}
\centering
\caption{The elemental abundances of the PNe, ordered by HASH ID, have been calculated by {\scshape neat}. Log(X/H) + 12 was used to represent the abundances. The elemental abundances for the bulge, disk, and halo are taken from KH22, while the solar abundances are sourced from A19. In this study, the mean values are calculated as the average of all PNe abundances. The number of references is shown in brackets.}
    \begin{tabular}{ccccccccc}
    \hline\hline
    Ref. & Source &	He/H	&	N/H	&	O/H	&	Ne/H	&	S/H	&	Cl/H	&	Ar/H	\\
    \hline
A19 & Solar	&	10.93	&	7.83	&	8.69	&	7.93	&	7.12	&	5.5	&	6.4	\\
   \hline
   Tan24 & Bulge (124)	&	11.06	$\pm$	0.08	&	8.27	$\pm$	0.30	&	8.70	$\pm$	0.15	&	8.14	$\pm$	0.18	&	6.90	$\pm$	0.22	&	6.44	$\pm$	0.23	&	5.14	$\pm$	0.26	\\
   \hline
KH22 & Bulge	&	11.04	&	8.2	&	8.63	&	7.93	&	6.91	&	5.77	&	6.32	\\
& Disk	&	11.04	&	8.13	&	8.61	&	7.97	&	6.8	&	5.05	&	6.31	\\
& Halo (13)	&	11.02	&	7.85	&	8.09	&	7.46	&	6.34	&	3.84	&	5.71	\\
\hline
This Work	&	Bulge (248)	&	11.32	$\pm$	0.04	&	7.80	$\pm$	0.08	&	8.50	$\pm$	0.07	&	7.55	$\pm$	0.08	&	6.42	$\pm$	0.11	&	5.69	$\pm$	0.10	&	6.57	$\pm$	0.07	\\
	&	Thin Disk (756)	&	11.32	$\pm$	0.04	&	7.93	$\pm$	0.08	&	8.47	$\pm$	0.07	&	7.81	$\pm$	0.09	&	6.53	$\pm$	0.11	&	5.91	$\pm$	0.10	&	6.62	$\pm$	0.08	\\
	&	Thick Disk (365)	&	11.22	$\pm$	0.04	&	7.65	$\pm$	0.08	&	8.41	$\pm$	0.05	&	7.72	$\pm$	0.08	&	6.43	$\pm$	0.10	&	5.67	$\pm$	0.08	&	6.43	$\pm$	0.07	\\
	&	Halo (80)	&	11.17	$\pm$	0.03	&	7.66	$\pm$	0.08	&	8.40	$\pm$	0.05	&	7.79	$\pm$	0.08	&	6.20	$\pm$	0.10	&	5.55	$\pm$	0.07	&	6.24	$\pm$	0.07	\\
	&	Milky Way (1,449)	&	11.26	$\pm$	0.04	&	7.76	$\pm$	0.08	&	8.44	$\pm$	0.06	&	7.72	$\pm$	0.08	&	6.47	$\pm$	0.07	&	6.39	$\pm$	0.10	&	5.71	$\pm$	0.09	\\
\hline 

\hline
HASH ID & PN G \\
\hline
11	&	000.1-01.1	&	11.29	$\pm$	0.03	&	6.93	$\pm$	0.05	&	8.75	$\pm$	0.02	&	-	&	5.80	$\pm$	0.09	&	6.23	$\pm$	0.08	&	7.12	$\pm$	0.04	\\
13	&	000.1-05.6	&	11.52	$\pm$	0.02	&	8.39	$\pm$	0.11	&	8.82	$\pm$	0.25	&	-	&	7.04	$\pm$	0.16	&	-	&	6.87	$\pm$	0.13	\\
14	&	000.1+17.2	&	11.58	$\pm$	0.05	&	8.31	$\pm$	0.02	&	8.38	$\pm$	0.05	&	-	&	6.61	$\pm$	0.06	&	-	&	-	\\
17	&	000.1-02.3	&	11.28	$\pm$	0.03	&	8.26	$\pm$	0.16	&	10.18	$\pm$	0.36	&	-	&	6.71	$\pm$	0.12	&	-	&	7.48	$\pm$	0.17	\\
19	&	000.2-01.9	&	11.75	$\pm$	0.05	&	8.65	$\pm$	0.02	&	8.41	$\pm$	0.01	&	-	&	6.97	$\pm$	0.03	&	-	&	7.17	$\pm$	0.03	\\
20	&	000.2-04.6	&	11.13	$\pm$	0.01	&	8.22	$\pm$	0.03	&	8.79	$\pm$	0.11	&	7.79	$\pm$	0.10	&	6.49	$\pm$	0.04	&	5.73	$\pm$	0.11	&	6.80	$\pm$	0.06	\\
22	&	000.3-04.6	&	11.31	$\pm$	0.07	&	7.40	$\pm$	0.11	&	7.57	$\pm$	0.23	&	-	&	5.77	$\pm$	0.08	&	-	&	5.90	$\pm$	0.12	\\
23	&	000.3+12.2	&	11.22	$\pm$	0.01	&	8.69	$\pm$	0.05	&	9.11	$\pm$	0.08	&	-	&	7.27	$\pm$	0.07	&	5.47	$\pm$	0.09	&	6.71	$\pm$	0.06	\\
25	&	000.3-02.8	&	10.92	$\pm$	0.04	&	5.86	$\pm$	0.03	&	6.75	$\pm$	0.01	&	-	&	4.34	$\pm$	0.09	&	-	&	5.30	$\pm$	0.04	\\
26	&	000.4-01.9	&	-	&	7.64	$\pm$	0.10	&	-	&	-	&	6.09	$\pm$	0.65	&	6.15	$\pm$	0.35	&	-	\\
27	&	000.4-02.9	&	11.09	$\pm$	0.03	&	8.59	$\pm$	0.11	&	8.80	$\pm$	0.13	&	-	&	6.79	$\pm$	0.21	&	-	&	6.29	$\pm$	0.08	\\
29	&	000.5+01.9	&	11.37	$\pm$	0.08	&	7.63	$\pm$	0.12	&	7.82	$\pm$	0.11	&	-	&	6.54	$\pm$	0.14	&	-	&	6.43	$\pm$	0.08	\\
30	&	000.5-03.1	&	10.43	$\pm$	0.04	&	8.87	$\pm$	0.07	&	9.45	$\pm$	0.13	&	8.06	$\pm$	0.09	&	7.22	$\pm$	0.09	&	5.42	$\pm$	0.09	&	5.91	$\pm$	0.08	\\
\dots&	\dots 	& \dots 	& \dots 	& \dots 	& \dots 	& \dots 	& \dots 	& \dots  	\\
    \hline
    
    \hline
    \end{tabular}
    \label{T:Abundance}
\end{table*}

\onecolumn
\small
\begin{longtable}{ccccccc}
\caption{The least-squares fit parameters for the element abundance ratios in Fig. \ref{F:Abundance_plot} and \ref{F:A1} are given for the Galactic components and the Milky Way as a whole.}
\label{T:Abundance_R}\\
\hline
Abundance Ratio&	Ref.  &Region & y-Intercept &  Slope & Correlation Coefficient & Number of Objects \\
\hline
\endfirsthead

\multicolumn{7}{c}%
{{\bfseries \tablename\ \thetable{} -- continued}} \\
\hline
Abundance Ratio&	Ref.  &Region & y-Intercept &  Slope & Correlation Coefficient & Object \\
\hline
\endhead

\hline
\endfoot

\hline \hline
\endlastfoot

\multirow{5}{*}\textbf{	Cl/He	}&This Work&	Bulge	&	-4.99	$\pm$	0.13	&	0.94	$\pm$	0.42	&	0.42	&	126	\\
	&&	Thin Disk	&	-6.56	$\pm$	0.07	&	1.10	$\pm$	0.21	&	0.51	&	395	\\
	&&	Thick Disk	&	-7.41	$\pm$	0.10	&	1.17	$\pm$	0.32	&	0.52	&	183	\\
	&&	Halo	&	1.27	$\pm$	0.23	&	0.38	$\pm$	0.87	&	0.16	&	40	\\
	&&	Milky Way	&	-6.35	$\pm$	0.05	&	1.07	$\pm$	0.16	&	0.50	&	744	\\
    \hline
\multirow{5}{*}\textbf{	O/N	}&This Work&	 Bulge	&	4.60	$\pm$	0.10	&	0.50	$\pm$	0.12	&	0.59	&	185	\\
	&&	Thin Disk	&	3.88	$\pm$	0.06	&	0.58	$\pm$	0.07	&	0.60	&	587	\\
	&&	Thick Disk	&	5.03	$\pm$	0.08	&	0.44	$\pm$	0.09	&	0.60	&	232	\\
	&&	Halo	&	4.83	$\pm$	0.19	&	0.47	$\pm$	0.23	&	0.61	&	40	\\
	&&	Milky Way	&	4.39	$\pm$	0.04	&	0.52	$\pm$	0.05	&	0.59	&	1044	\\
    \hline	
\multirow{5}{*}\textbf{	Ne/N	}&This Work&	 Bulge	&	5.79	$\pm$	0.23	&	0.23	$\pm$	0.26	&	0.42	&	21	\\
	&&	Thin Disk	&	5.18	$\pm$	0.11	&	0.32	$\pm$	0.16	&	0.40	&	114	\\
	&&	Thick Disk	&	4.66	$\pm$	0.12	&	0.38	$\pm$	0.12	&	0.69	&	61	\\
	&&	Halo	&	2.01	$\pm$	0.35	&	0.73	$\pm$	0.38	&	0.84	&	10	\\
	&&	Milky Way	&	4.84	$\pm$	0.08	&	0.36	$\pm$	0.09	&	0.54	&	206	\\
    \hline	
\multirow{5}{*}\textbf{	S/N	}&This Work&	 Bulge	&	-0.02	$\pm$	0.05	&	0.82	$\pm$	0.07	&	0.91	&	178	\\
	&&	Thin Disk	&	-0.25	$\pm$	0.04	&	0.85	$\pm$	0.05	&	0.86	&	509	\\
	&&	Thick Disk	&	0.13	$\pm$	0.06	&	0.81	$\pm$	0.08	&	0.87	&	182	\\
	&&	Halo	&	0.26	$\pm$	0.17	&	0.78	$\pm$	0.21	&	0.83	&	37	\\
	&&	Milky Way	&	-0.08	$\pm$	0.03	&	0.83	$\pm$	0.04	&	0.87	&	906	\\
    \hline	
\multirow{5}{*}\textbf{	Ar/N	}&This Work&	 Bulge	&	3.47	$\pm$	0.09	&	0.40	$\pm$	0.10	&	0.56	&	177	\\
	&&	Thin Disk	&	2.80	$\pm$	0.06	&	0.48	$\pm$	0.07	&	0.57	&	516	\\
	&&	Thick Disk	&	4.20	$\pm$	0.08	&	0.29	$\pm$	0.09	&	0.48	&	203	\\
	&&	Halo	&	4.49	$\pm$	0.16	&	0.22	$\pm$	0.21	&	0.39	&	36	\\
	&&	Milky Way	&	3.29	$\pm$	0.04	&	0.42	$\pm$	0.05	&	0.55	&	932	\\
    \hline	
\multirow{5}{*}\textbf{	Ne/O	}&This Work&	 Bulge	&	1.42	$\pm$	0.16	&	0.71	$\pm$	0.34	&	0.68	&	29	\\
	&&	Thin Disk	&	2.59	$\pm$	0.10	&	0.61	$\pm$	0.18	&	0.53	&	152	\\
	&&	Thick Disk	&	-0.88	$\pm$	0.10	&	1.02	$\pm$	0.17	&	0.82	&	97	\\
	&&	Halo	&	1.04	$\pm$	0.18	&	0.81	$\pm$	0.31	&	0.78	&	25	\\
	&&	Milky Way	&	1.31	$\pm$	0.07	&	0.76	$\pm$	0.12	&	0.66	&	303	\\
    \cline{2-7}
	&KH22&	Milky Way	&	–0.06	$\pm$	0.72	&	0.94	$\pm$	0.08	&	0.72	&	116	\\
	&Tan24&	Bulge	&	0.90	$\pm$	0.05	&	0.32	$\pm$	0.46	&	0.86	&	105	\\
    \hline
\multirow{5}{*}\textbf{	S/O	}&This Work&	 Bulge	&	0.95	$\pm$	0.10	&	0.64	$\pm$	0.14	&	0.61	&	183	\\
	&&	Thin Disk	&	1.48	$\pm$	0.06	&	0.60	$\pm$	0.07	&	0.62	&	558	\\
	&&	Thick Disk	&	1.29	$\pm$	0.10	&	0.60	$\pm$	0.15	&	0.53	&	221	\\
	&&	Halo	&	1.53	$\pm$	0.22	&	0.56	$\pm$	0.37	&	0.46	&	48	\\
	&&	Milky Way	&	1.35	$\pm$	0.04	&	0.60	$\pm$	0.06	&	0.59	&	1010	\\
    \cline{2-7}
	&KH22&	Milky Way	&	0.024	$\pm$	1.17	&	0.77	$\pm$	0.14	&	0.47	&	117	\\
    \hline	
\multirow{5}{*}\textbf{	Ar/O	}&This Work&	 Bulge	&	1.53	$\pm$	0.07	&	0.59	$\pm$	0.10	&	0.70	&	217	\\
	&&	Thin Disk	&	1.04	$\pm$	0.04	&	0.66	$\pm$	0.06	&	0.73	&	640	\\
	&&	Thick Disk	&	2.16	$\pm$	0.06	&	0.51	$\pm$	0.09	&	0.61	&	300	\\
	&&	Halo	&	3.21	$\pm$	0.11	&	0.36	$\pm$	0.21	&	0.46	&	60	\\
	&&	Milky Way	&	1.35	$\pm$	0.03	&	0.61	$\pm$	0.04	&	0.69	&	1217	\\
    \cline{2-7}
	&KH22&	Milky Way	&	–1.45	$\pm$	0.83	&	0.91 	$\pm$	0.10	&	0.66	&	117	\\
	&Tan24&	Bulge	&	0.95	$\pm$	0.06	&	-1.86	$\pm$	0.48	&	0.84	&	122	\\
    \hline	
\multirow{5}{*}\textbf{	Ar/S	}&This Work&	 Bulge	&	3.55	$\pm$	0.09	&	0.47	$\pm$	0.12	&	0.58	&	173	\\
	&&	Thin Disk	&	2.75	$\pm$	0.05	&	0.59	$\pm$	0.07	&	0.66	&	497	\\
	&&	Thick Disk	&	4.64	$\pm$	0.08	&	0.29	$\pm$	0.11	&	0.41	&	193	\\
	&&	Halo	&	4.25	$\pm$	0.14	&	0.31	$\pm$	0.19	&	0.52	&	41	\\
	&&	Milky Way	&	3.30	$\pm$	0.04	&	0.50	$\pm$	0.05	&	0.60	&	904	\\
    \hline	
\multirow{5}{*}\textbf{	Ar/Cl	}&This Work&	 Bulge	&	3.89	$\pm$	0.10	&	0.48	$\pm$	0.14	&	0.58	&	124	\\
	&&	Thin Disk	&	3.58	$\pm$	0.06	&	0.53	$\pm$	0.09	&	0.58	&	368	\\
	&&	Thick Disk	&	4.61	$\pm$	0.08	&	0.32	$\pm$	0.11	&	0.46	&	165	\\
	&&	Halo	&	4.78	$\pm$	0.13	&	0.27	$\pm$	0.22	&	0.43	&	36	\\
	&&	Milky Way	&	3.84	$\pm$	0.04	&	0.48	$\pm$	0.06	&	0.56	&	693	\\
    \hline	
\multirow{5}{*}\textbf{	Ar/He	}&This Work&	 Bulge	&	-2.67	$\pm$	0.08	&	0.82	$\pm$	0.24	&	0.47	&	216	\\
	&&	Thin Disk	&	-5.13	$\pm$	0.05	&	1.04	$\pm$	0.15	&	0.54	&	606	\\
	&&	Thick Disk	&	0.45	$\pm$	0.07	&	0.53	$\pm$	0.20	&	0.34	&	288	\\
	&&	Halo	&	0.16	$\pm$	0.12	&	0.55	$\pm$	0.50	&	0.32	&	58	\\
	&&	Milky Way	&	-3.47	$\pm$	0.04	&	0.89	$\pm$	0.11	&	0.49	&	1168	\\
    \hline	
\multirow{5}{*}\textbf{	Cl/O	}&This Work&	 Bulge	&	3.03	$\pm$	0.14	&	0.31	$\pm$	0.20	&	0.30	&	130	\\
	&&	Thin Disk	&	3.19	$\pm$	0.08	&	0.32	$\pm$	0.10	&	0.33	&	414	\\
	&&	Thick Disk	&	3.73	$\pm$	0.12	&	0.23	$\pm$	0.19	&	0.20	&	186	\\
	&&	Halo	&	3.93	$\pm$	0.23	&	0.19	$\pm$	0.52	&	0.13	&	42	\\
	&&	Milky Way	&	3.15	$\pm$	0.06	&	0.31	$\pm$	0.08	&	0.30	&	772	\\
    \cline{2-7}
	&KH22&	Milky Way	&	–2.26	$\pm$	1.03	&	0.83	$\pm$	0.12	&	0.56	&	107	\\
	&Tan24&	Bulge	&	0.76	$\pm$	0.10	&	1.44	$\pm$	0.83	&	0.59	&	119	\\
    \hline
\multirow{5}{*}\textbf{	S/Ne	}&This Work&	 Bulge	&	-0.18	$\pm$	0.38	&	0.89	$\pm$	0.81	&	0.48	&	23	\\
	&&	Thin Disk	&	4.23	$\pm$	0.15	&	0.34	$\pm$	0.24	&	0.28	&	127	\\
	&&	Thick Disk	&	3.36	$\pm$	0.20	&	0.42	$\pm$	0.31	&	0.35	&	72	\\
	&&	Halo	&	-1.21	$\pm$	0.32	&	0.97	$\pm$	0.44	&	0.82	&	15	\\
	&&	Milky Way	&	3.24	$\pm$	0.11	&	0.45	$\pm$	0.18	&	0.35	&	237	\\
    \hline	
\multirow{5}{*}\textbf{	Cl/Ne	}&This Work&	 Bulge	&	2.62	$\pm$	0.28	&	0.38	$\pm$	0.56	&	0.35	&	20	\\
	&&	Thin Disk	&	2.86	$\pm$	0.13	&	0.36	$\pm$	0.23	&	0.32	&	112	\\
	&&	Thick Disk	&	3.39	$\pm$	0.18	&	0.26	$\pm$	0.38	&	0.19	&	68	\\
	&&	Halo	&	-0.41	$\pm$	0.27	&	0.72	$\pm$	0.49	&	0.67	&	16	\\
	&&	Milky Way	&	2.72	$\pm$	0.09	&	0.36	$\pm$	0.18	&	0.31	&	216	\\
    \hline	
\multirow{5}{*}\textbf{	Ar/Ne	}&This Work&	 Bulge	&	5.51	$\pm$	0.23	&	0.15	$\pm$	0.46	&	0.14	&	28	\\
	&&	Thin Disk	&	3.57	$\pm$	0.09	&	0.40	$\pm$	0.14	&	0.48	&	139	\\
	&&	Thick Disk	&	2.92	$\pm$	0.10	&	0.45	$\pm$	0.15	&	0.59	&	91	\\
	&&	Halo	&	1.81	$\pm$	0.20	&	0.54	$\pm$	0.30	&	0.73	&	18	\\
	&&	Milky Way	&	3.29	$\pm$	0.07	&	0.42	$\pm$	0.11	&	0.48	&	276	\\
    \hline	
\multirow{5}{*}\textbf{	Cl/S	}&This Work&	 Bulge	&	3.78	$\pm$	0.13	&	0.29	$\pm$	0.18	&	0.34	&	110	\\
	&&	Thin Disk	&	3.73	$\pm$	0.08	&	0.31	$\pm$	0.11	&	0.34	&	334	\\
	&&	Thick Disk	&	4.87	$\pm$	0.12	&	0.10	$\pm$	0.16	&	0.13	&	137	\\
	&&	Halo	&	3.95	$\pm$	0.27	&	0.23	$\pm$	0.40	&	0.24	&	29	\\
	&&	Milky Way	&	3.90	$\pm$	0.06	&	0.27	$\pm$	0.08	&	0.31	&	610	\\
    \hline	
\end{longtable}
\normalsize
\twocolumn


\appendix
\setcounter{figure}{0} \renewcommand{\thefigure}{A.\arabic{figure}}

\begin{figure*}
\begin{center}
\includegraphics[width=0.8\textwidth]{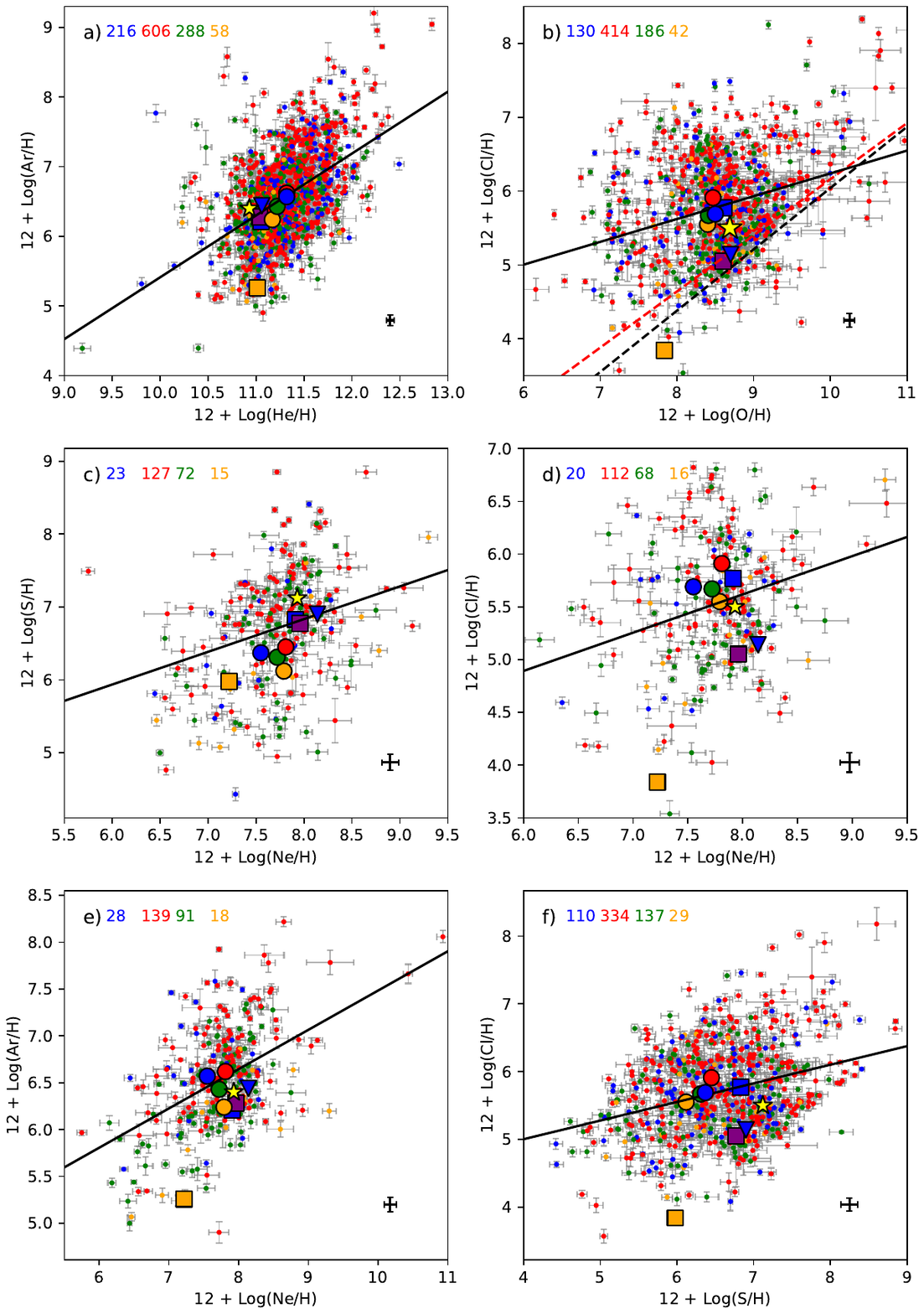}
\caption{The elemental abundances of the PNe plotted on a 12 + log(X/H) scale for correlation coefficients ranging from 0.3 to 0.5. The descriptions are same with given in Fig. \ref{F:Abundance_plot}.}
\label{F:A1}
\end{center}
\end{figure*}

\bsp	
\label{lastpage}
\end{document}
